\documentclass[%
 aip,
 amsmath,amssymb,
preprint,%
author-year,%
]{revtex4-1}

\usepackage{graphicx}
\usepackage{dcolumn}
\usepackage{bm}
\usepackage{subcaption}
\usepackage{color}
\usepackage{upgreek}

\usepackage[utf8]{inputenc}
\usepackage[T1]{fontenc}
\usepackage{mathptmx}
\usepackage{hyperref}

\begin{document}

\preprint{AIP/123-QED}

\title[COVID-19 Urban Bus]{On the Utility of a Well-Mixed Model for Predicting Disease Transmission on an Urban Bus}

\author{Zhihang Zhang}
 \affiliation{%
Department of Naval Architecture and Marine Engineering, University of Michigan
}%
 \author{Jesse Capecelatro}%
  \altaffiliation[Also at ]{Department of Aerospace Engineering, University of Michigan}
 \affiliation{%
Department of Mechanical Engineering, University of Michigan
}%

\author{Kevin Maki}
 \email{kjmaki@umich.edu.}
  \affiliation{%
Department of Naval Architecture and Marine Engineering, University of Michigan
}%
%

\date{\today}

\begin{abstract}
The transport of virus-laden aerosols from a host to a susceptible person is governed by complex turbulent airflow, and physics related to breathing, coughing and sneezing, mechanical and passive ventilation, thermal buoyancy effects, surface deposition, masks, and air filtration.  In this paper, we study the infection risk via airborne transmission on an urban bus using unsteady Reynolds-Averaged Navier--Stokes equations and a passive-scalar model of the virus-laden aerosol concentration. Results from these simulations are directly compared to the widely-used well mixed model, and show significant differences in the concentration field and number of inhaled particles.  Specifically, in the limit of low mechanical ventilation rate, the well-mixed model will over-predict concentration far from the infected passenger, and substantially underpredict concentration near the infected passenger. The results reported herein also apply to other enclosed spaces.


\end{abstract}

\maketitle

\section{\label{sec:intro}Introduction}
Accurately predicting the transmission of airborne diseases is critical for assessing the associated risk to individuals and to inform decision making with regard to societal controls. For example, it is now recognized that airborne transmission is the dominant transmission mode of SARS-CoV-2~\citep{Zhang20}, which is responsible for the COVID-19 pandemic. The majority of so-called `super-spreading' events during the COVID-19 pandemic has taken place in indoor settings with poor ventilation~\citep[see, e.g.,][]{setti2020airborne,shen2020community,morawska2020can,Miller20}. The precise infectious dose an individual would be exposed to in a given setting depends on the turbulent nature of the airflow that carries the virus from one person to another and the various factors that contribute to the underlying fluid dynamics.

Due to its simplicity, the well-mixed model~\citep{Riley78, Miller20} remains the predominant tool to predict transmission for risk assessment. It is based primarily on the assumption that the concentration in an enclosed space is spatially uniform such that an ordinary-differential equation can be written as a balance of the generation rate due to infected persons, and the destruction due to causes such as natural decay of infectivity, removal due to ventilation, filtration, or deposition, etc.  
 \cite{Miller20,Bazant21} use the well-mixed model to analyze transmission for indoor spaces, and include data from super spreader events that have occurred throughout the COVID-19 pandemic. However, indoor spaces are rarely well mixed, and developing simple models for predicting exposure of virus-laden aerosols under realistic settings remains challenging. 

Computational fluid dynamics (CFD) is increasingly being used to study the detailed flow physics of airborne diseases. \cite{Talaat21} use steady Reynolds-average Navier--Stokes (RANS) simulations to predict the flow field in the cabin of a Boeing 737 airplane to assess mitigation and safety of air travel. Several recent papers presented detailed analyses of the flow physics of a cough using direct numerical simulation~\citep{fabregat2021direct,monroe2021role,liu2021peering}. \cite{Dbouk21, Dbouk20} use Eulerian--Lagrangian particle tracking to study the physics of coughing, effectiveness of face masks, and the transmission risk in an elevator.  While detailed studies using RANS or large-eddy simulation (LES) are able to predict the complex flow and pathways of virus-laden aerosols, it is not practical to simulate all indoor spaces and airflow or occupancy arrangements.  

\citet{Lau21} refined the well-mixed mode by adding advection and diffusion terms to the equation governing the concentration. The challenge with this approach is how to effectively model advection, since it is inherently nonlinear in an Eulerian setting, and depends on many parameters, including specific room geometry, people and their movement within the enclosed space, and effects of mechanical and natural ventilation.  \cite{Guo21} combined the well-mixed model and the Wells--Riley equation with CFD simulations to  assess the need to control spread in hospital spaces.  They found that the well-mixed model, which does not consider spatial variability, fails to accurately predict risk, and does not provide the spatial details that are needed to space people within a given room or enclosed zone.




This paper uses detailed simulations that model turbulence and the geometry of the enclosed space to predict the transport of virus-laden aerosols throughout the passenger cabin of an urban bus. The mechanical ventilation system on the bus is used to create two different flow scenarios, one in which forced convection and turbulence dominate the aerosol transport, and another in which the ventilation system plays a much smaller role such that diffusion and natural convection play a more important role.  The detailed numerical results are compared with estimates from a well-mixed model, and it is shown that while this model is instructive for general guidance, it ignores the spatial dependence of aerosols and this limits its accuracy.

\section{Modeling of Aerosol Transport}
In this paper, we consider the smallest aerosols only, those with diameter less than $10$ $\upmu$m.  Under this assumption, aerosols are represented with the continuum field $C=C(\mathbf{x},t)$, which is transported passively with the unsteady RANS (URANS) velocity field~\cite{Zhang21}. Details on both the well-mixed model and CFD simulations are summarized herein.

\paragraph{Well-Mixed Model} As described above, a common approach to modeling the aerosol in a confined space employs the well-mixed assumption in which the concentration field does not vary in space.  It is not clear what is required for aerosols to be perfectly mixed, yet this assumption is widely used since the resulting model is so simple. The well mixed concentration field $\tilde{C}=\tilde{C}(t)$ satisfies the model equation
\begin{equation}
\frac{{\rm d} \tilde{C}}{{\rm d} t} = \frac{\lambda}{V} - \gamma\tilde{C},
\end{equation}
where $V$ is the volume of the enclosed space, $\gamma$ is the loss-rate coefficient that accounts for the decay, deposition, loss due to ventilation, filtration, etc. Here, the tilde denotes a spatially averaged quantity. The source of aerosols is represented by the emission rate $\lambda$. The solution for the well-mixed concentration is simply
\begin{equation}\label{eqn:c}
\tilde{C} = \frac{\lambda}{V\gamma}\left(1 - e^{-\gamma t}\right).
\end{equation}

The equilibrium value at large $t$ of the concentration is $\tilde{C}_{\infty}=\lambda/V\gamma$, where 
 $\gamma^{-1}$ is the time scale for the spatially uniform concentration to reach the equilibrium value.  The relationship between the spatial average and local concentration is
 
 \begin{equation}
 \tilde{C} = \frac{1}{V}\int_V C(\mathbf{x},t)\,{\rm d}V,
 \end{equation}
which is the link between the CFD simulations and the well-mixed model prediction.

\paragraph{Emission of respiratory droplets}

An important quantity in any airborne transmission and risk study is the rate at which the host produces virus-laden aerosols.  \cite{vuorinen2020modelling, Miller20, Mittal20b,Bazant21} all present a range of virus shedding rate from an infected individual.   The emission rate is defined as the number of virus-laden aerosols per time as $\lambda$ s$^{-1}$.  It depends on the stage of infection, masking, and breathing rate, and this quantity is difficult to accurately assess.  For modeling in CFD, we use a continuous breathing model due to the fact that the breathing period is of order seconds and the time horizon of the analysis is minutes.  In this work, the shedding rate is assumed to be $\lambda = 50$~s$^{-1}$, and breathing rate $\dot{V}_b = 0.1$~l-s$^{-1}$. The breathing rate depends strongly on the activity of the person, and can range from 6-100 l-min$^{-1}$~\citep{Mittal20b}.

\paragraph{Susceptible Person Disease Contraction}

The contraction of the disease by a susceptible person via the airborne route requires inhalation of the virus-laden aerosols.  The minimum number of aerosols that are required for one to become ill is not clearly understood.  The range is thought to be between 100 and 1000 \citep{vuorinen2020modelling}, and methods like the Wells--Riley~\citep{Riley78} (which uses a well-mixed assumption), or dose-response functions~\citep{Bazant21} describe the probability of infection.  In this work, the minimum infective dose (MID) of $N_{{\rm b, crit}}\approx50$ is used based on the analysis of 20 COVID-19 spreading events~\citep{Kolinski20}. Also, the Wells--Riley approach is used when comparing the CFD results with the well-mixed model. The Wells-Riley assumes that the concentration is uniform throughout space in the enclosed volume, and the value of the concentration is the long-time equilibrium value from the well-mixed model.  The probability of inhaling an infectious dose is calculated as~\citep[Eqn.~1]{Riley78}: 
\begin{equation}\label{eqn:WR}
P = 1 - e^{rt},
\end{equation}
where $r = \tilde{C}_{\infty}\dot{V_b}$.   The present CFD results will give direct insight into the validity of the underlying assumptions used in the Wells--Riley equation.

The number of inhaled aerosols, $N_b$, is related to the concentration field as 
\begin{equation}\label{eqn:n}
N_b(\mathbf{x},t) = \int C(\mathbf{x},t) \dot{V_b} {\rm d}t.
\end{equation}
If the well-mixed model is used, the concentration and number of inhaled particles is no longer a function of space, and the integral can be solved directly to determine the number of aerosols inhaled as 
\begin{equation}\label{eqn:n-well-mixed}
N_b(t) = \frac{\dot{V_b}\lambda}{V\gamma}\left[ t + \frac{1}{\gamma}\left(e^{-\gamma t} - 1\right) \right].
\end{equation}
This equation shows that for long time $t\gamma\gg1$ the number of inhaled aerosols grows linearly.  On the other hand, for the initial time, that is $t\gamma\ll1$, the solution is proportional to $t^2$, \emph{i.e.}
\begin{equation}
N_b(t) = \frac{\dot{V}_b\lambda}{2V}t^2 \mbox{ for } t \gamma\ll1.
\label{eqn:aerosol-inhale}
\end{equation}
The quadratic nature of the initial period is acutely problematic and exposure for time up to $t\gamma \approx 1$ should be considered to be quadratic according to the well-mixed assumption. 

Equations~\eqref{eqn:c} and~\eqref{eqn:n-well-mixed} constitute the description of virus transport in indoor spaces that is used throughout the literature and policy making throughout the world.  In this paper, we compare the well-mixed formulation with the more accurate CFD prediction of the time and spatially varying concentration field in order to determine whether the well-mixed model is appropriate for accurately predicting risk of infection.

\paragraph{URANS-based Numerical Method}
In this work, the turbulent airflow inside an urban bus is determined using the URANS equations.  The effects of turbulence are represented with the $k-\epsilon$ turbulence model.  Buoyancy effects due to temperature differences are considered through a Buossinesq approximation.  Aerosols are modeled with a continuum approximation.  The number of inhaled aerosols is computed for all locations within the confined space through numerical solutions of Eq.~\eqref{eqn:n}. The method allows for details of the geometry of the enclosed space and for the aerodynamic effects of the ventilation system.  The customized solver is based on the OpenFOAM opensource CFD platform. The  concentration field $C$ is governed by the convection-diffusion equation:
\begin{equation} 
\frac{\partial{C}}{\partial{t}}+\nabla\cdot(\mathbf{u}C)-\nabla\cdot\left(D_{\rm eff}\nabla{C}\right)=0,
\end{equation}
where $D_{\rm eff}=\nu_{t}/{\rm Sc}_{t}+\nu/{\rm Sc}$, and ${\rm Sc}_{t} = {\rm Sc} = 1$ are the turbulent and laminar Schmidt numbers, respectively.  Additional details can be found in the paper \cite{Zhang21}. The custom solver is available on Github.\footnote{\url{https://github.com/zhihangz/covid-transmission}}

\section{Description of Bus and Numerical Setup}

The University of Michigan operates  54 urban busses to transport students, faculty, staff, and visitors across campus.  Each bus can hold up to 70 passengers (35 seated and the rest standing).  
Aerosol transport inside the bus is governed by the heating ventilation air conditions (HVAC) system, thermal effects due to HVAC and passengers, the opening of windows and doors, and passenger movement.  The well-mixed model accounts for HVAC and windows and doors.

\begin{figure}
	\begin{subfigure}{1\textwidth}
	\includegraphics[width=1.0\textwidth]{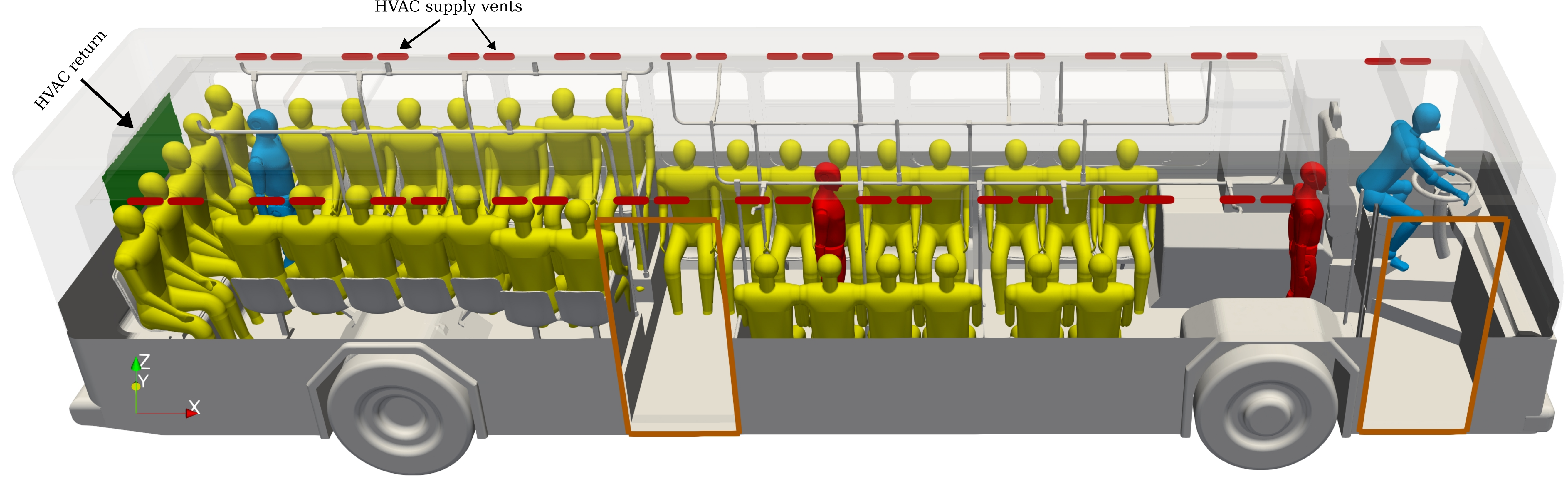}
	\caption{Bus interior with HVAC return and supply, and passenger arrangement.\label{fig:bus-interior}}
	\end{subfigure}
	\begin{subfigure}{1\textwidth}
	\includegraphics[width=1.0\textwidth]{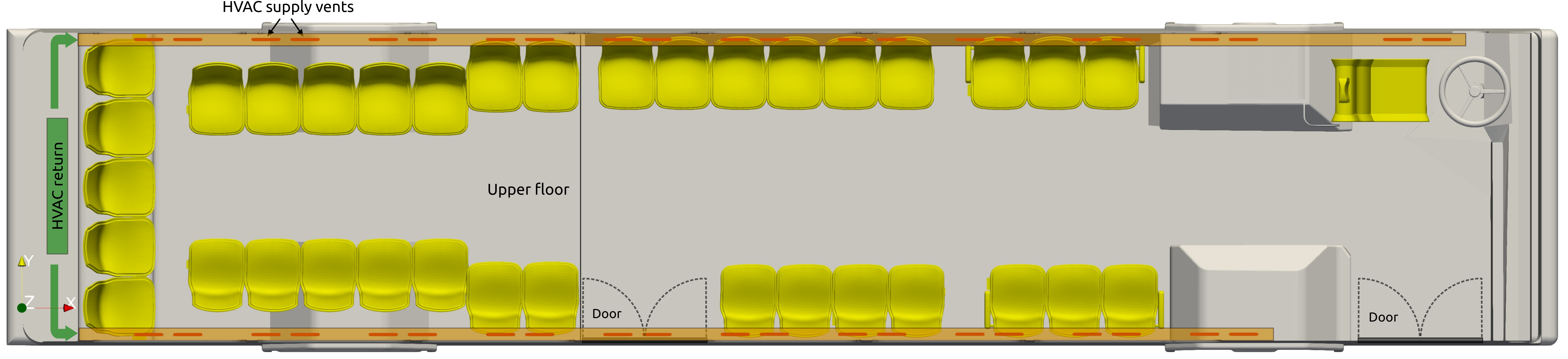}
	\caption{Top view of the bus interior.}
	\end{subfigure}
	\caption{Bus interior used in the CFD simulations.}
\end{figure}

The bus nominal dimensions are $12.1\times 2.58 \times 2.95$~m and the interior of the bus is approximately 2,000~ft$^3$ (56.6~m$^3$). The HVAC system consists of a single fan in the rear of the bus that gathers air from the interior through a single vent on the rear bulkhead, and distributes the conditioned air through a system of supply vents that are oriented downward along the header rails on both sides of the bus.  See Figure~\ref{fig:bus-interior} for a view inside the bus with the HVAC return vent and supply vents. The HVAC system can supply conditioned air at a maximal rate of 2,500~ft$^3$-min$^{-1}$ (70.8~m$^3$-min$^{-1}$), and the conditioned air includes 20\% that is added from outside. There are 42 supply vents, each with dimension of $9\times1$~in ($0.229\times 0.0254$~m), and the return vent is $4\times1.5$~ft ($1.22\times0.457$~m). 

The interior geometry of the bus is digitized with a laser scanner and used to generate a mesh that includes all of the interior geometry including seating, steps, and hand rails. Manikins are placed in the seats, and three standing passengers are located near the front, in the middle, and in the far rear of the bus. The standing passengers are used as hosts, and it is assumed that their emission rate is $\lambda=50$~s$^{-1}$.   The continuous breathing model is used, and the exhalation rate is $\dot{V}_b=0.1$~l-s$^{-1}$.  

A mesh refinement study was performed and the converged results on a mesh of 5.87 million unstructured cells are used in the present analysis.  The largest cells in the domain are cubes of edge length 125~mm, and the smallest cells on the surfaces such as the mouths of the passengers and the HVAC supply vents are 4 and 2~mm, respectively.

The turbulence intensity and length scale for the mouth of the passengers is 10\% and 7.5~mm. The intensity and length scale for the HVAC supply vents is 2.5\% and 5~mm. The temperature of the conditioned air is 20$^\circ$C, and the oral temperature of 37$^\circ$C is applied to the exhaled breath.

The boundary conditions on the HVAC supply vents enforce a downward velocity set by the fan flow-rate, and the aerosol concentration assumes an even redistribution of that which exits the passenger cabin through the HVAC return vents, with 20\% replaced by fresh uncontaminated air.

Four simulations are performed corresponding to two settings of the HVAC system, and two locations of the infected passenger.    The bus HVAC system when set at the maximal rate generates a strong current from front to rear in the bus, whereas when at ten percent of the maximum it is much less dominant.  The single HVAC return vent at the rear of the bus drives a net rearward flow in the passenger cabin, and to analyze the role of the HVAC system the passengers are organized into two groups corresponding to either in front or behind the infected host. Table~\ref{tab:run-matrix} summarizes the four simulations.

For each HVAC rate, a precursor simulation of length three minutes is run to establish equilibrium airflow conditions in the bus before the infected passenger begins to exhale virus-laden aerosols. 

\begin{table}
\caption{\label{tab:run-matrix}Simulation Parameters }
\begin{ruledtabular}
\begin{tabular}{lcr}
Run & Host Position & HVAC Rate\\
\hline
1 & Forward & 100\%\\
2 & Middle & 100\%\\
3 & Forward & 10\%\\
4 &  Middle & 10\%\\
\end{tabular}
\end{ruledtabular}
\end{table}

\section{Results}

\subsection{Flow Field}

\begin{figure}
	\begin{subfigure}{1\textwidth}
	\includegraphics[width=1.0\textwidth]{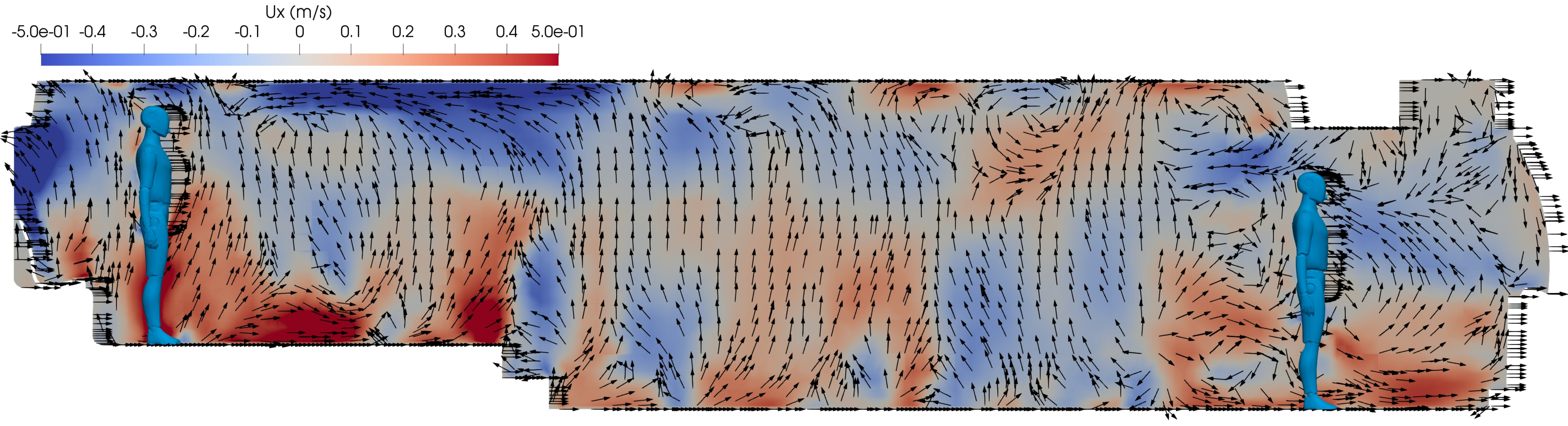}
	\caption{Run 1: HVAC at maximum rate.}
	\end{subfigure}
	\begin{subfigure}{1\textwidth}
	\includegraphics[width=1.0\textwidth]{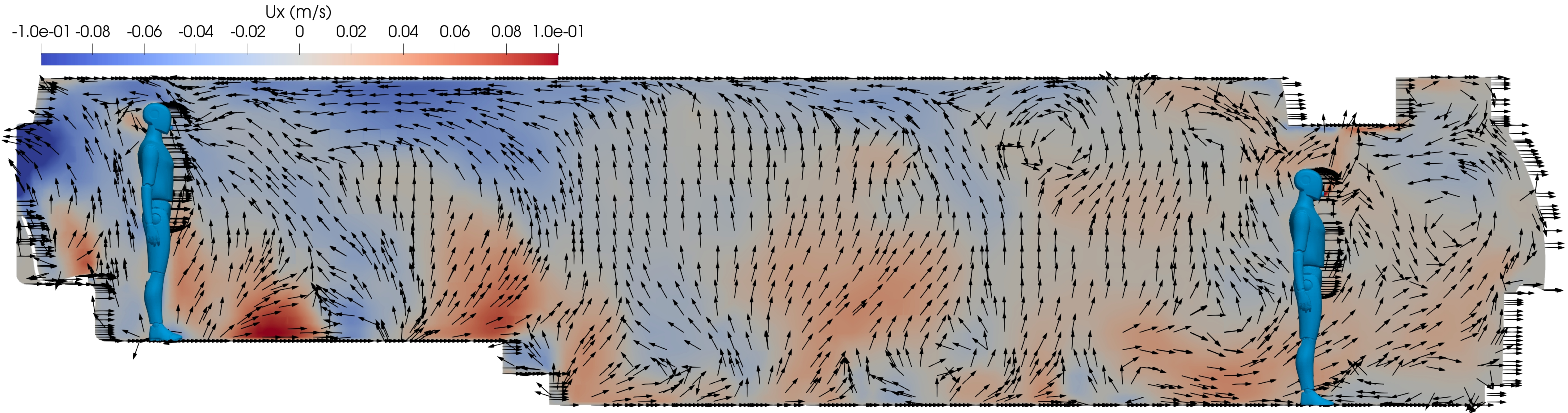}
	\caption{Run 3: HVAC at 10\% of maximum rate.}
	\end{subfigure}
	\caption{Instantaneous velocity vectors of mean $x$ velocity on center plane of the bus at $t$~=~15 min.} 
	\label{flow-field}
\end{figure}

The transport of the virus from host to susceptible via the airborne route depends on the turbulent fluid velocity field that is driven by breathing, ventilation, and thermal effects.   Two HVAC settings are investigated, and Fig.~\ref{flow-field} shows the instantaneous velocity vector fields, with color contours of the mean velocity in the $x$ direction, where $x$ points from the front to the rear of the bus.  Upward and downward currents are observed for both HVAC rates that indicate the airflow should effectively mix the concentration in the passenger cabin.  In addition,  for both cases the rearward velocity increases as one moves closer to the HVAC return at the back of the bus, but this effect is much stronger for the higher HVAC rate.  This is an important effect on how the HVAC system influences the transmission of the disease and the number of infected passengers. Another interesting phenomenon is the appearance of local circulation in front of the host for the lower HVAC rate while rearward currents present in the case with the higher HVAC rate. This causes the aerosols to be trapped in the front of the cabin in Run 3, which can be observed in the contours of inhaled aerosols.

\subsection{Aerosol Concentration}

\begin{figure}[!ht]
	\begin{subfigure}{0.418\textwidth}
	\includegraphics[width=\linewidth]{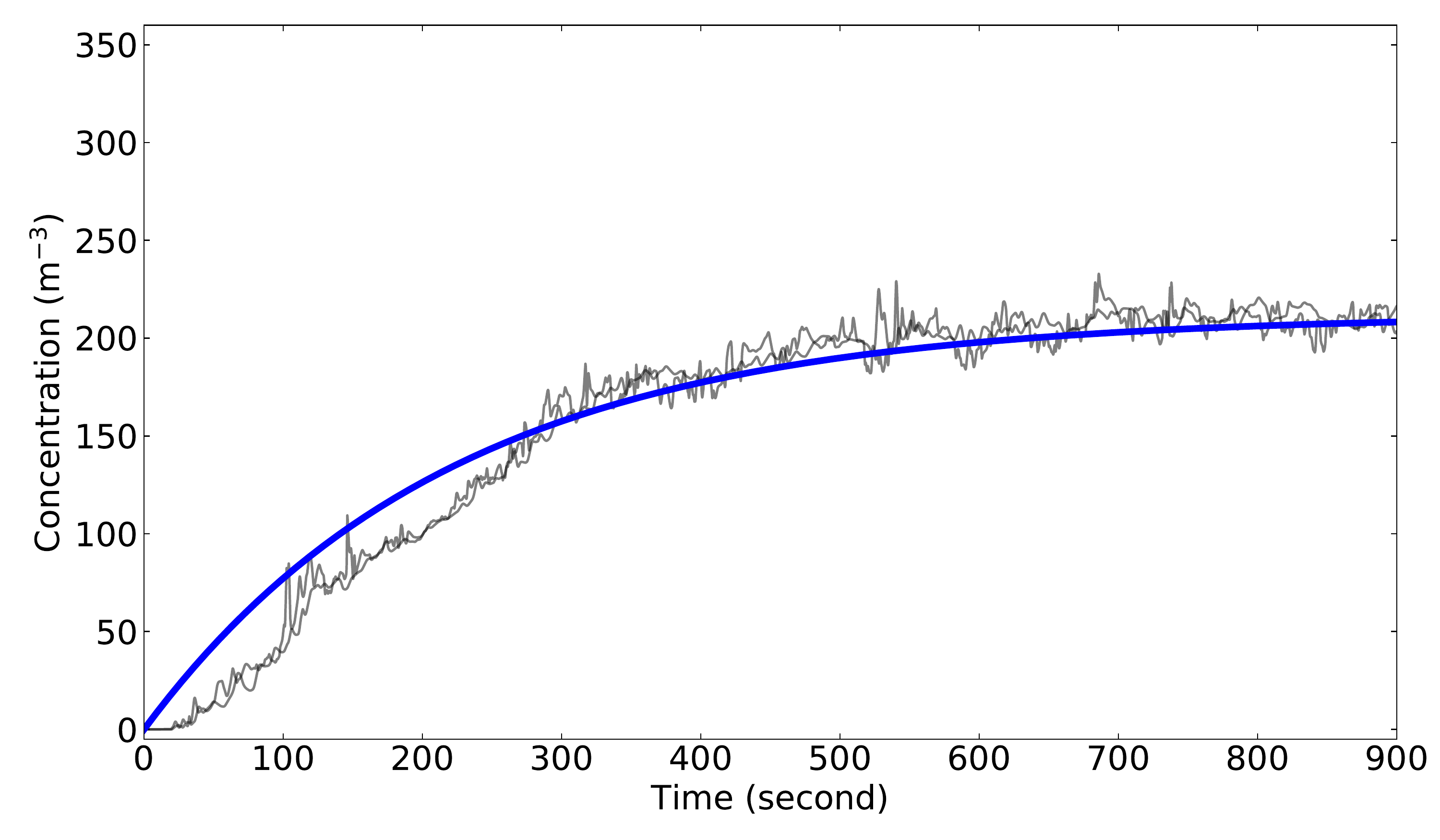}
	\caption{Run 1: in front of the host}
	\end{subfigure}
	\hfill
	\begin{subfigure}{0.418\textwidth}
	\includegraphics[width=\linewidth]{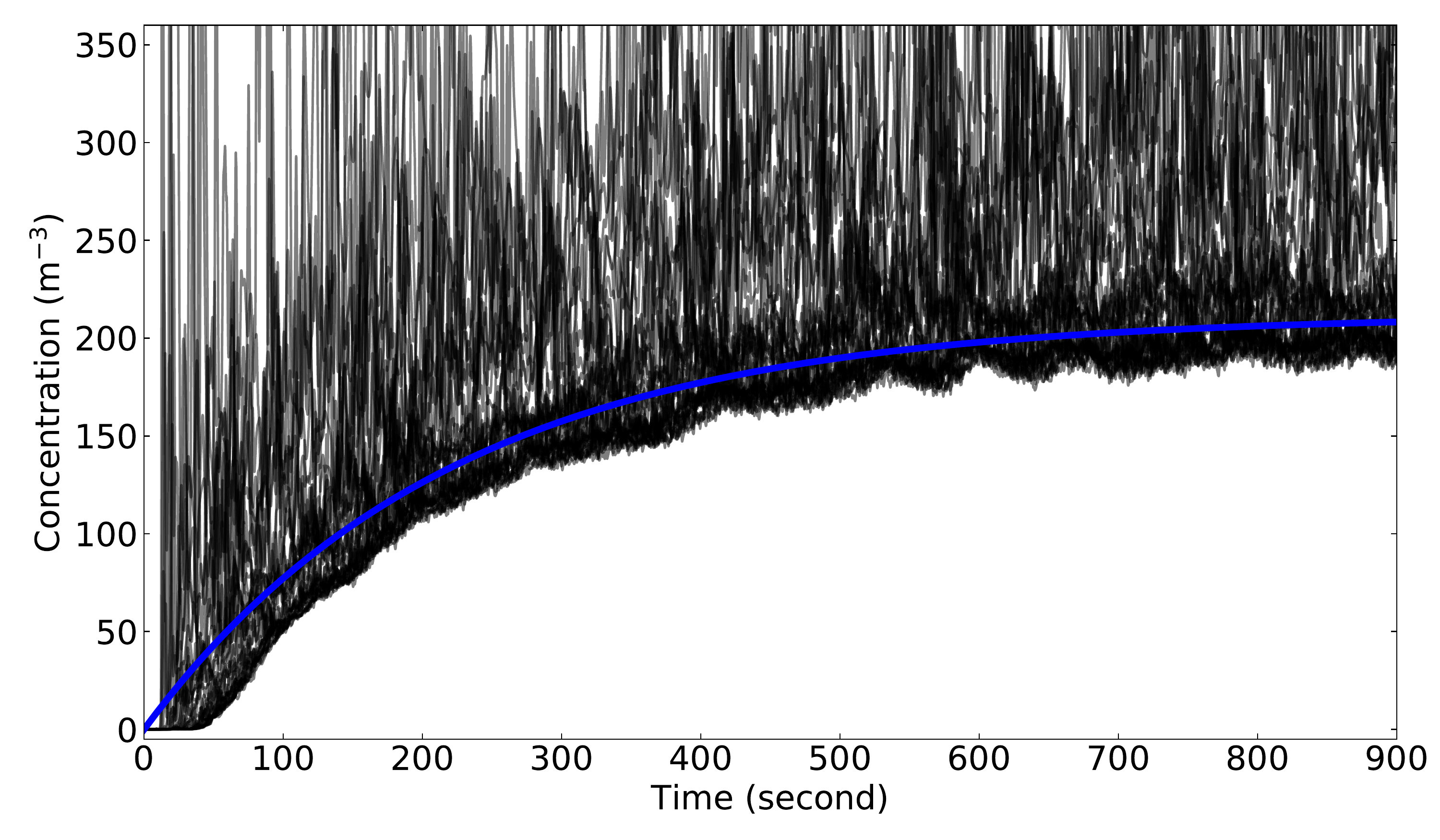}
	\caption{Run 1: behind the host}
	\end{subfigure}

	\begin{subfigure}{0.418\textwidth}
	\includegraphics[width=\linewidth]{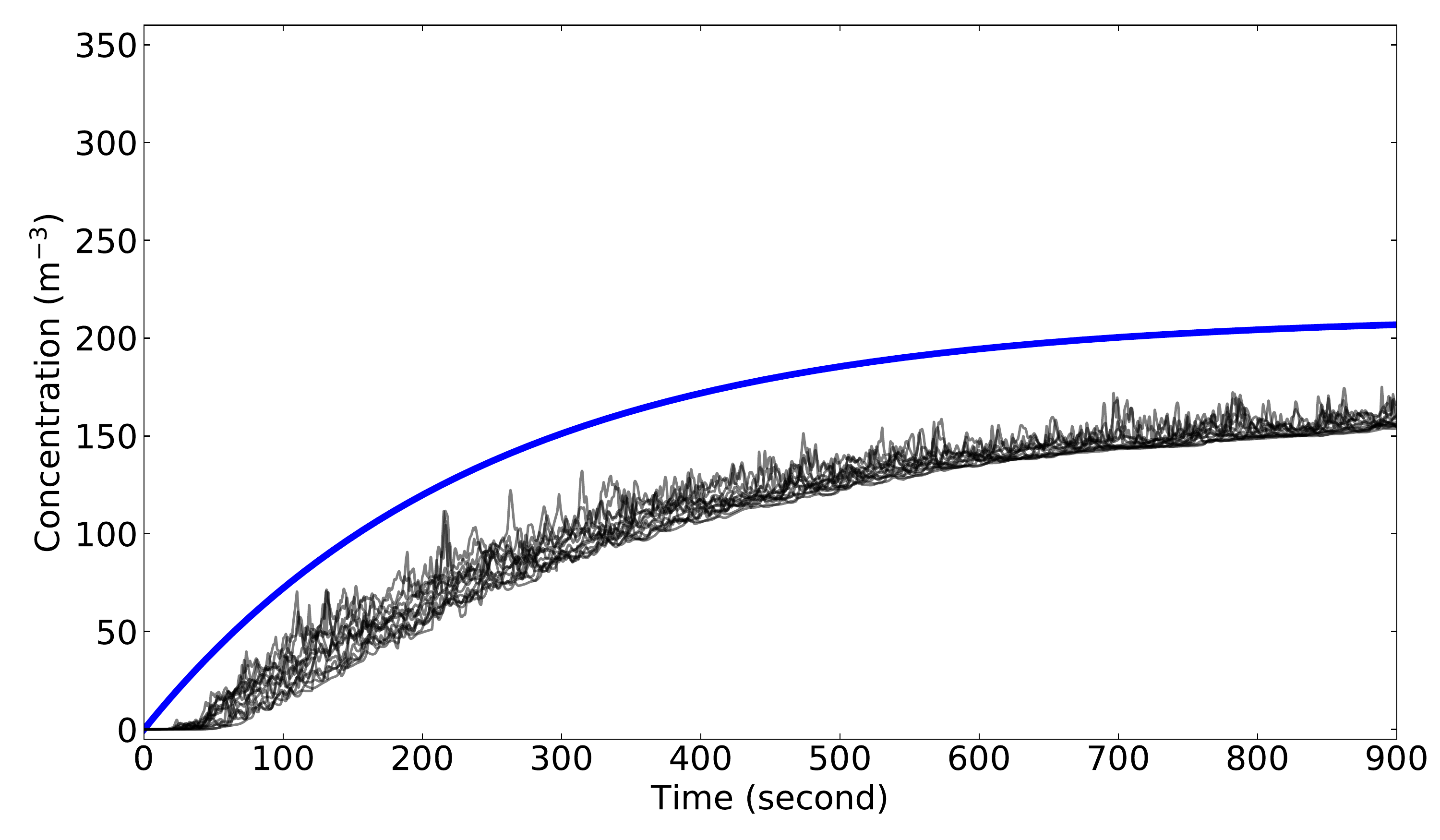}
	\caption{Run 2: in front of the host}
	\end{subfigure}
	\hfill
	\begin{subfigure}{0.418\textwidth}
	\includegraphics[width=\linewidth]{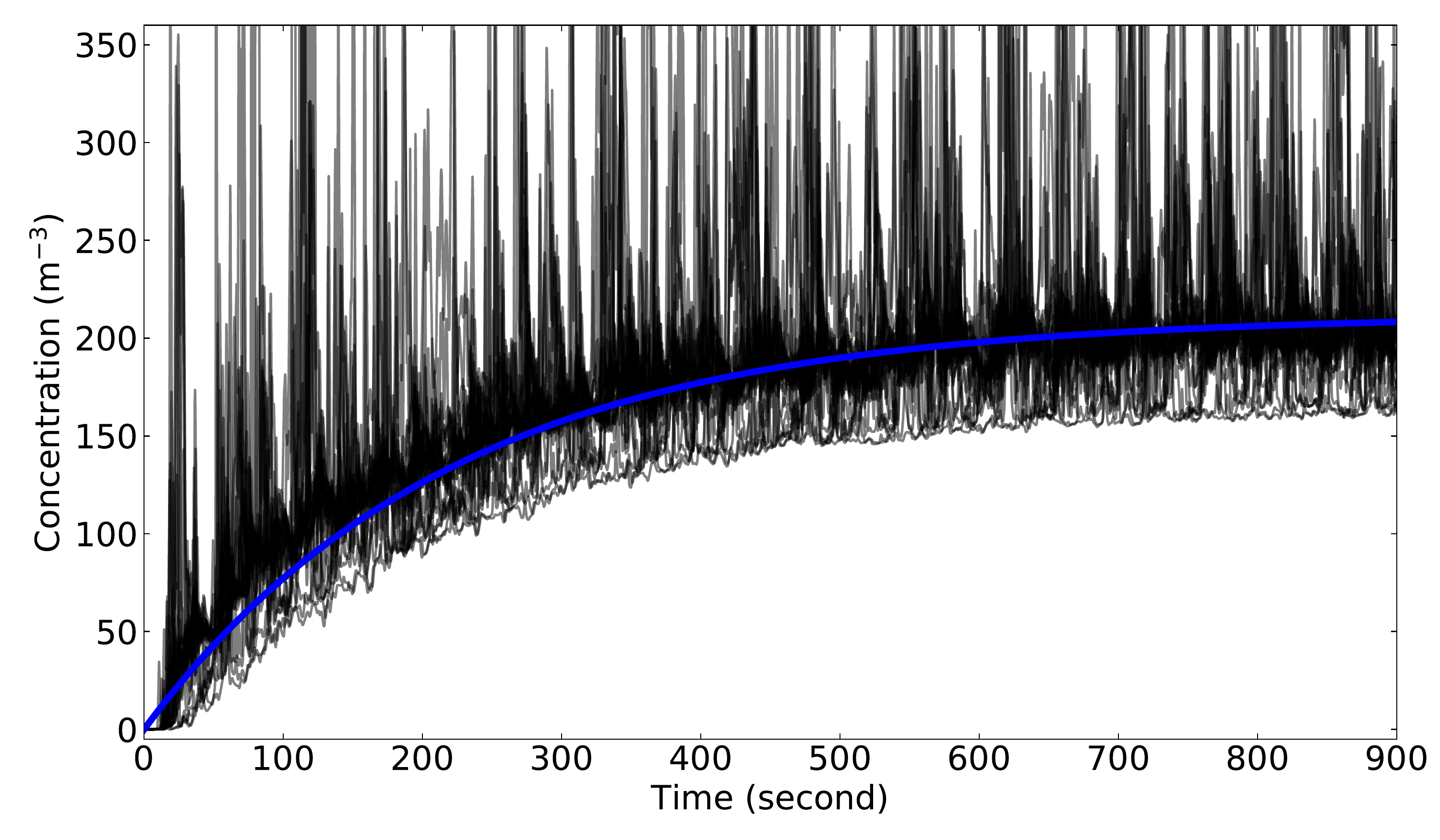}
	\caption{Run 2: behind the host}
	\end{subfigure}
	
	\begin{subfigure}{0.418\textwidth}
	\includegraphics[width=\linewidth]{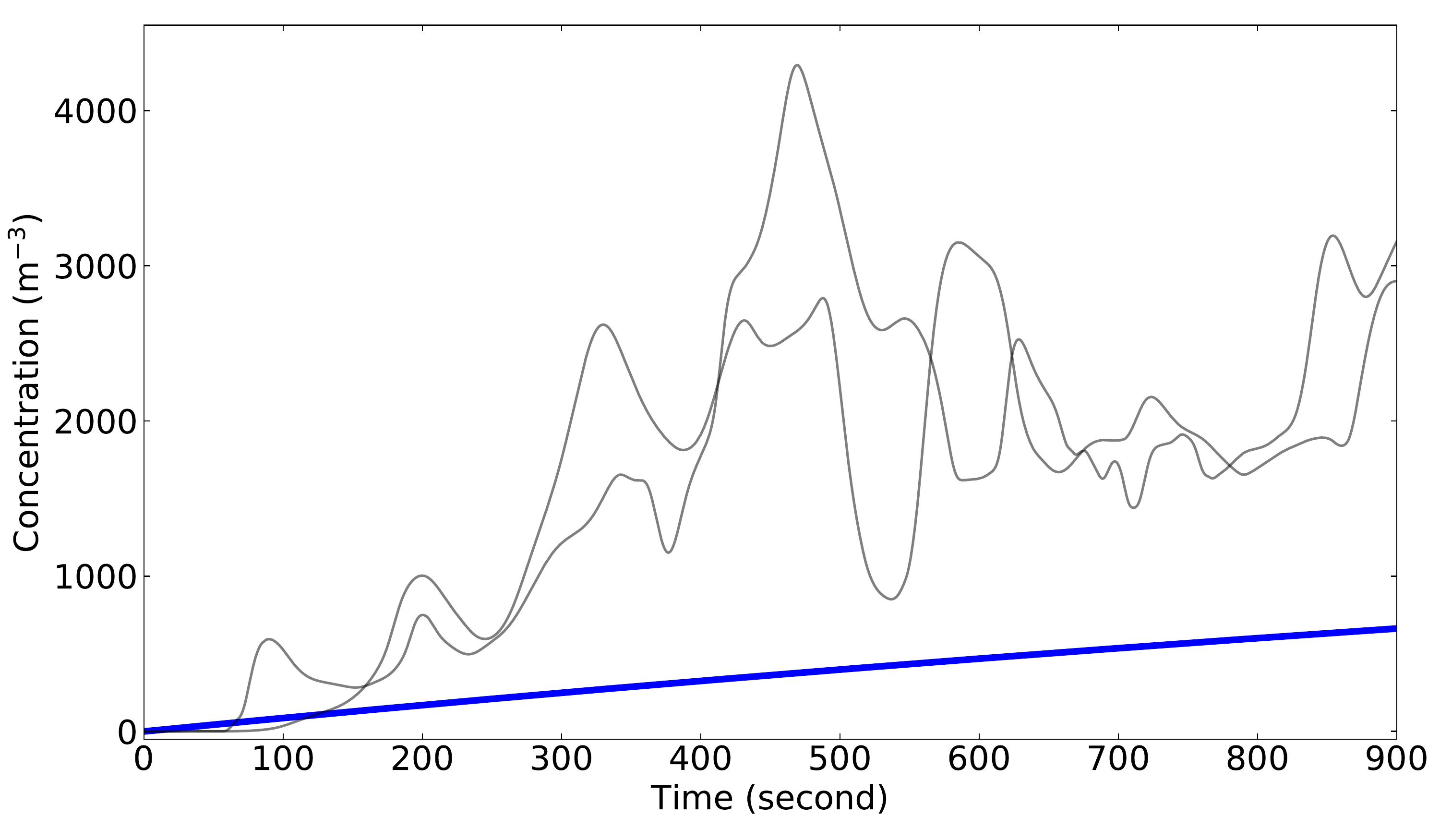}
	\caption{Run 3: in front of the host}
	\end{subfigure}
	\hfill
	\begin{subfigure}{0.418\textwidth}
	\includegraphics[width=\linewidth]{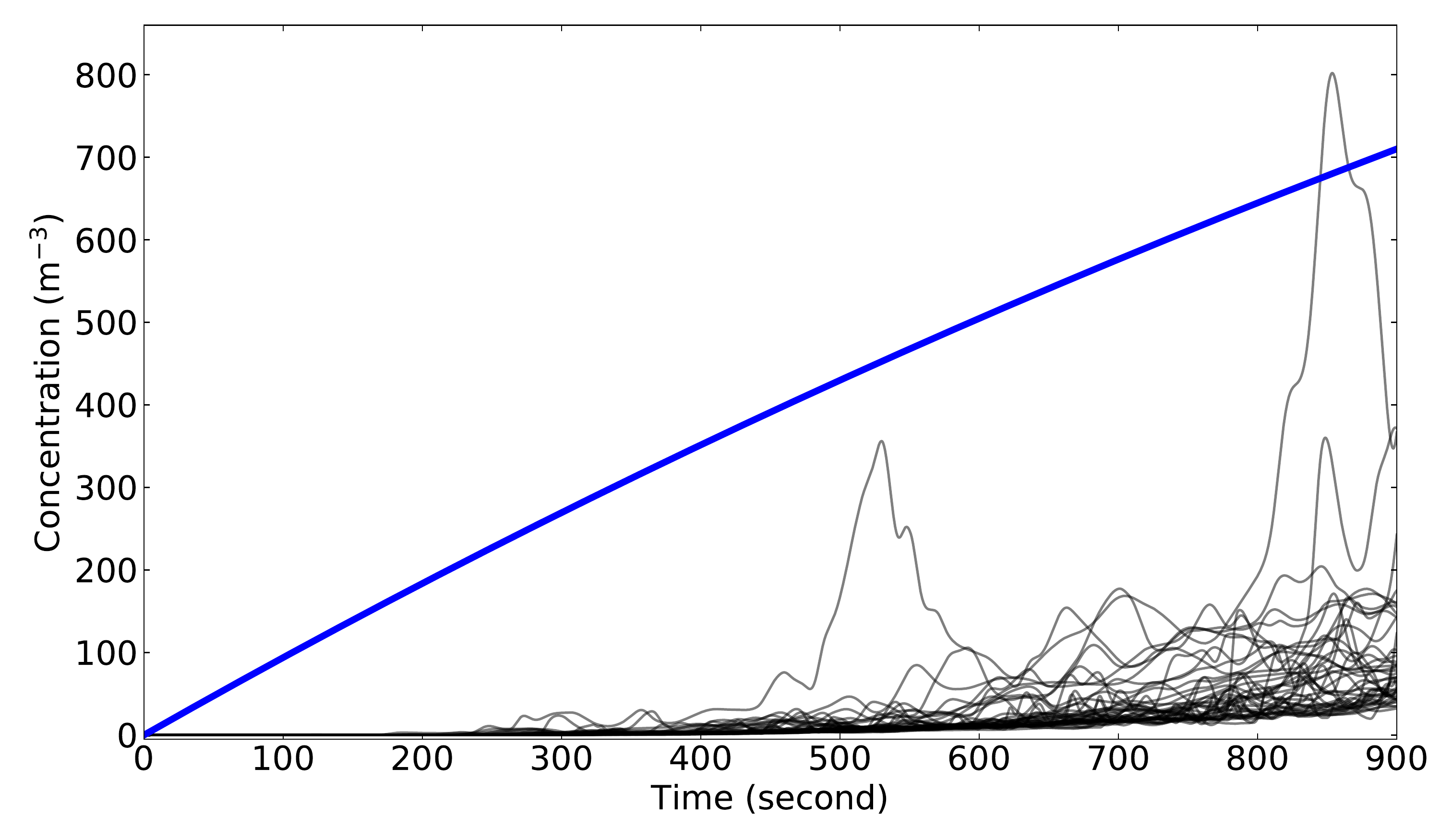}
	\caption{Run 3: behind the host}
	\end{subfigure}

	\begin{subfigure}{0.418\textwidth}
	\includegraphics[width=\linewidth]{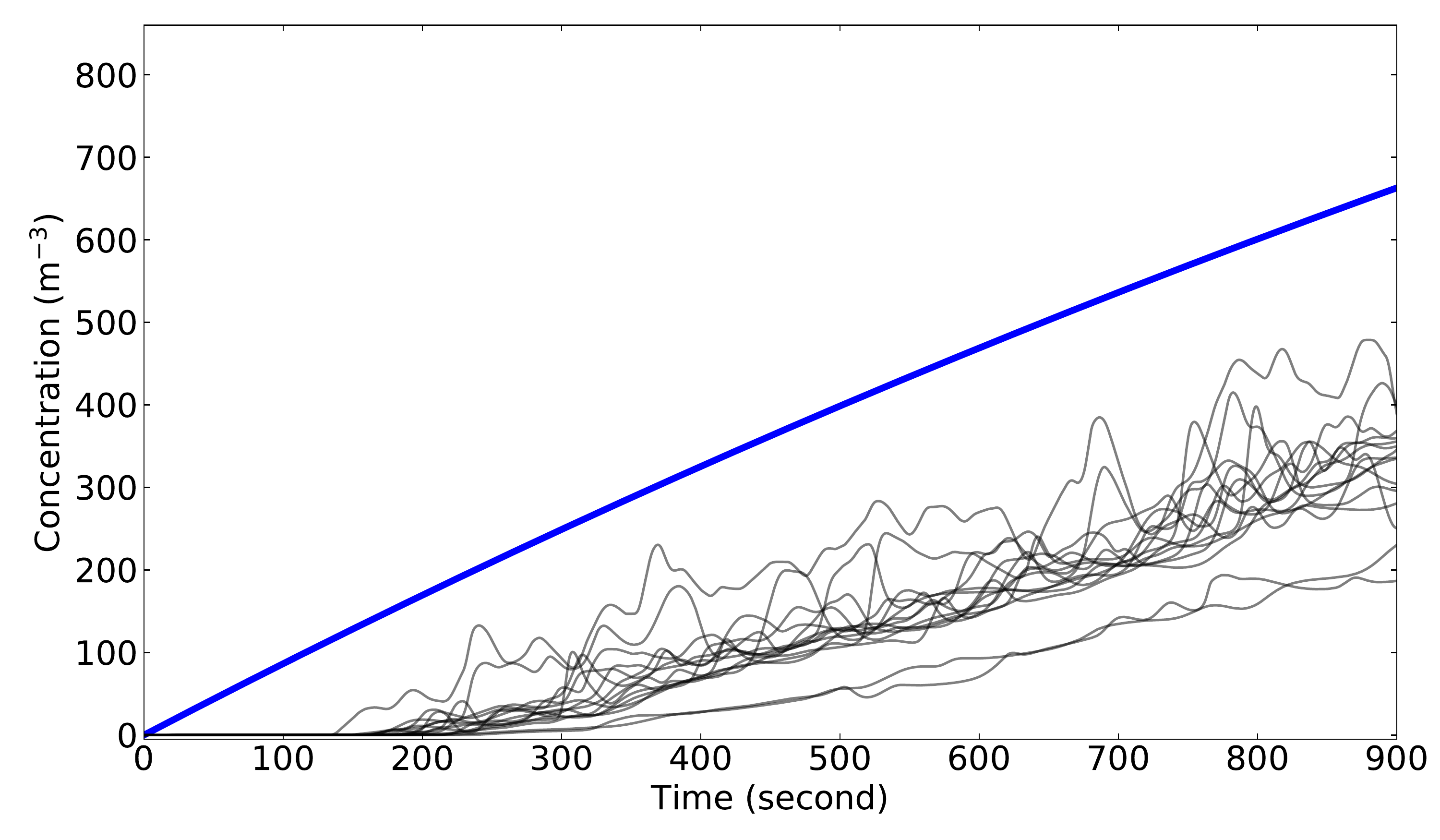}
	\caption{Run 4: in front of the host}
	\end{subfigure}
	\hfill
	\begin{subfigure}{0.418\textwidth}
	\includegraphics[width=\linewidth]{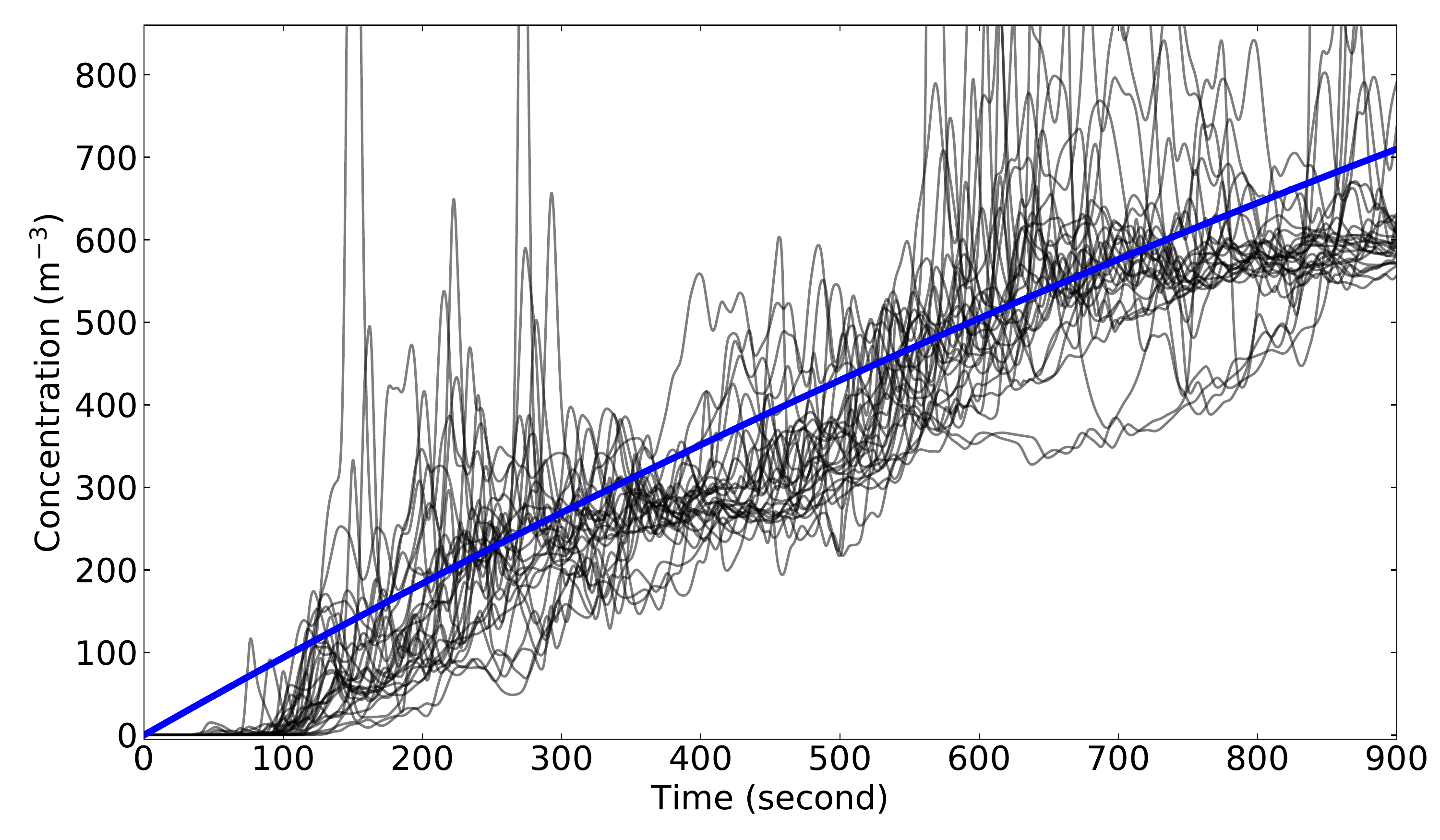}
	\caption{Run 4: behind the host}
	\end{subfigure}
	
		\caption{Time history of aerosol concentration at different passenger locations (black lines), compared with the well-mixed model (blue lines). \label{fig:conc}}

\end{figure}

The aerosol concentration at different probe locations is plotted as a function of time for each of the four simulations in Fig.~\ref{fig:conc}. There are a total of 42 probes representing 42 susceptible people, including 35 seated passenger, the driver, and 6 standing passengers through the centerline of the bus. The probes are all placed near the mouths.  The solution to the well-mixed model Eq.~\eqref{eqn:c} is plotted in the thick blue line. The loss-rate coefficient for the maximal HVAC setting is $\gamma = 4.17\times10^{-3}$ s$^{-1}$ ($\gamma^{-1} = 240$ s) and the  equilibrium concentration is $\tilde{C}_{\infty}=212$~m$^{-3}$ and for the low HVAC setting they are $\gamma = 4.17\times10^{-4}$ s$^{-1}$ ($\gamma^{-1} = 2398$ s), $\tilde{C}_{\infty}=2119$~m$^{-3}$. It is interesting to note that the general trend of the concentration follows that predicted by the well-mixed model for the highest HVAC rate, whereas the disagreement is significant for the low HVAC rate.  For Run~1, the concentration predicted by the CFD model exhibits large variability due to  turbulence. Also, there is a time-lag in the concentration field that corresponds to the time it takes for the first exhaled aerosols to reach the particular passenger.  In the well-mixed model this occurrs instantaneously.  The time shift depends on the distance from the host to susceptible passenger, and can be as large as $60$~s. This means that for a susceptible passenger that is far away from an host, it is only a matter of a minute until contaminated air is inhaled, regardless of the distance between the two passengers. Also note that the maximum concentration at any instant in time can be significantly greater than that predicted by the well-mixed model. Finally, the  concentration approaches equilibrium in the 15~min exposure time of the longest bus ride ($t\gamma\approx3.75$).

Similar observations can be made for Run~2 where the host is located in the middle of the bus.

The results for Run 3 with the low HVAC rate and the host forward show that the well-mixed model and the CFD predictions differ significantly (Figs.~\ref{fig:conc}e-f). For almost all passengers, the well-mixed model overpredicts the concentration, except for the two passengers in front of the host. For these two passengers, the concentration is significantly higher due to the low rearward flux from the HVAC system causing the concentration to accumulate in the front of the bus.  In the 15~min exposure, the concentration is still steadily increasing, and much larger than the equilibrium value for the high HVAC Runs 1 and 2.  

Run 4 is similar to 3, but the infected passenger is located in the middle of the bus. Figs.~\ref{fig:conc}g-h show the time histories of this run, and the agreement between the well-mixed model and the CFD is close.    Similar to runs 1 and 2, the instantaneous concentration can be much greater than that predicted by the well-mixed model. Also, the time lag for the low HVAC rate is increased to almost 100~s due to the reduced rearward flow induced by the HVAC fan.

The well-mixed model assumes that the concentration of the virus is instantaneously averaged in space.  The CFD computations yield the time-accurate evolution of the spatially evolving concentration field.  To analyze the way the virus spreads throughout the bus, the spatial mean and standard deviation are calculated according to
\begin{eqnarray}
\tilde{C}_{\rm CFD}(t) &=& \frac{1}{V}\int C(\mathbf{x},t) d\mathbf{x}\\
{\sigma}(t) &=&\left[ \frac{1}{V}\int \left(C(\mathbf{x},t) - \tilde{C}_{\rm CFD}\right)^2 d\mathbf{x} \right]^{1/2}.
\end{eqnarray}

\begin{figure}
    \centering
    \begin{subfigure}{0.48\textwidth}
        \includegraphics[width=\linewidth]{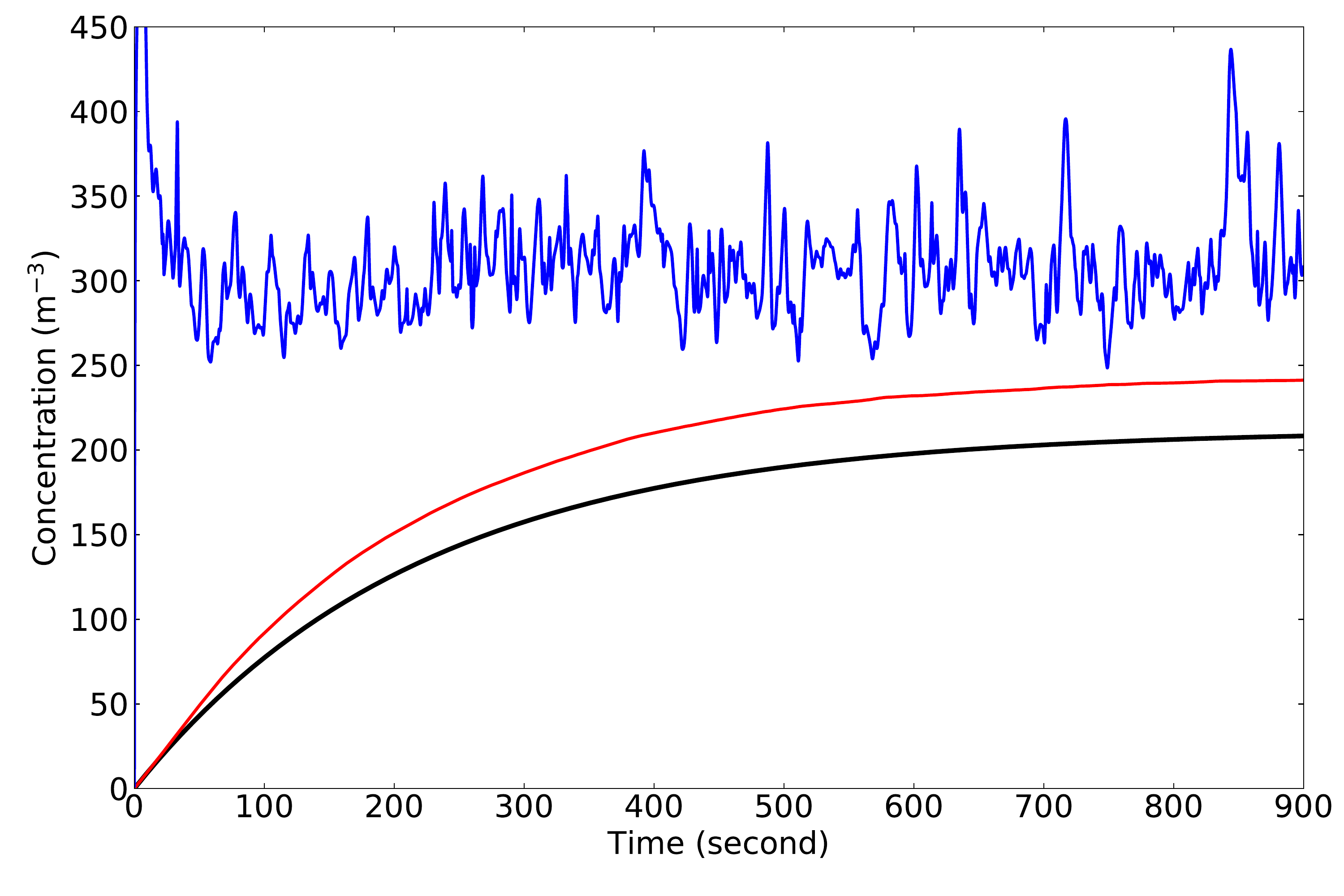}
        \caption{Run 1} \label{fig:average-run1}
    \end{subfigure}
    \hspace*{\fill}
    \begin{subfigure}{0.48\textwidth}
        \includegraphics[width=\linewidth]{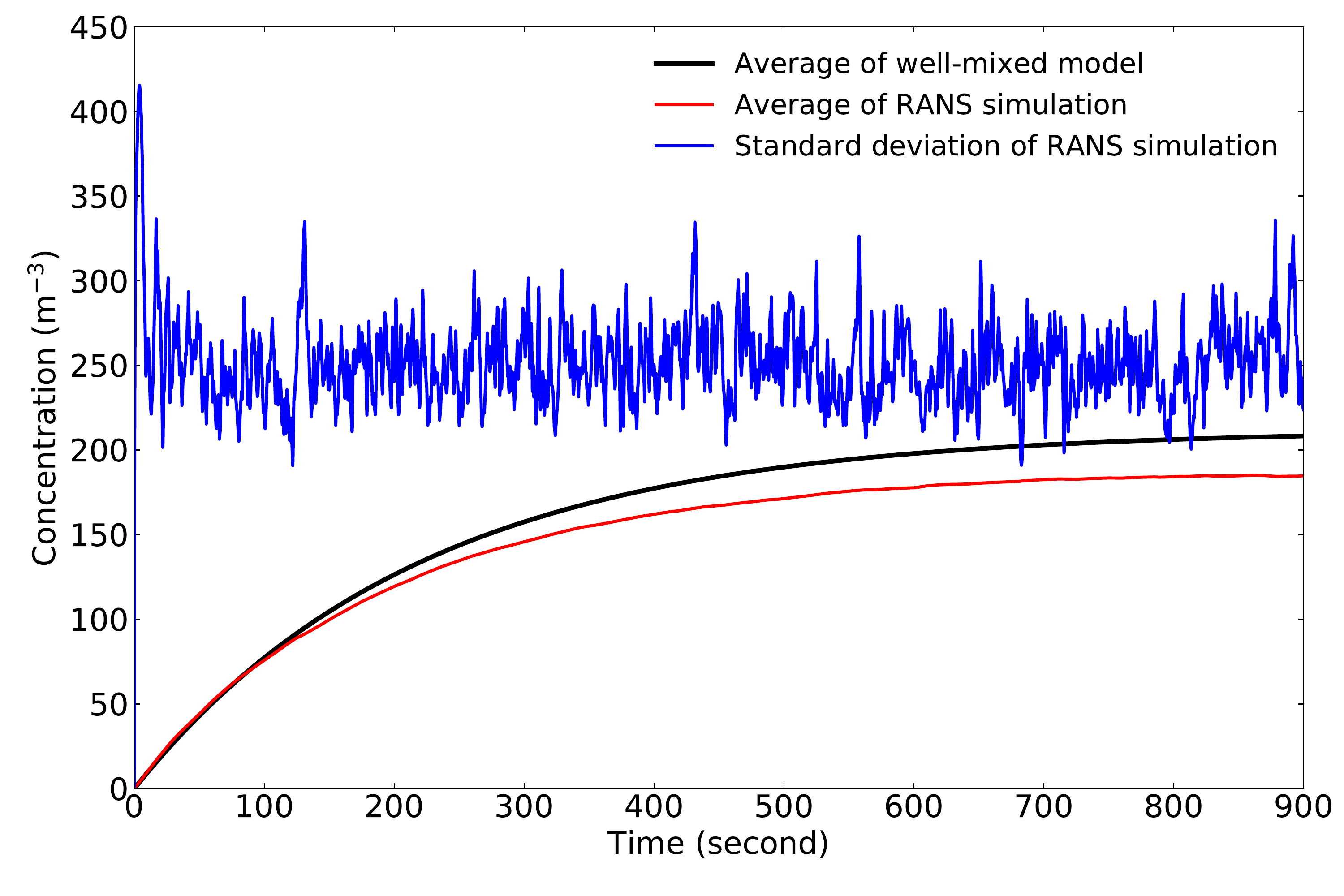}
        \caption{Run 2} \label{fig:average-run2}
    \end{subfigure}
    \centering
    \begin{subfigure}{0.48\textwidth}
        \includegraphics[width=\linewidth]{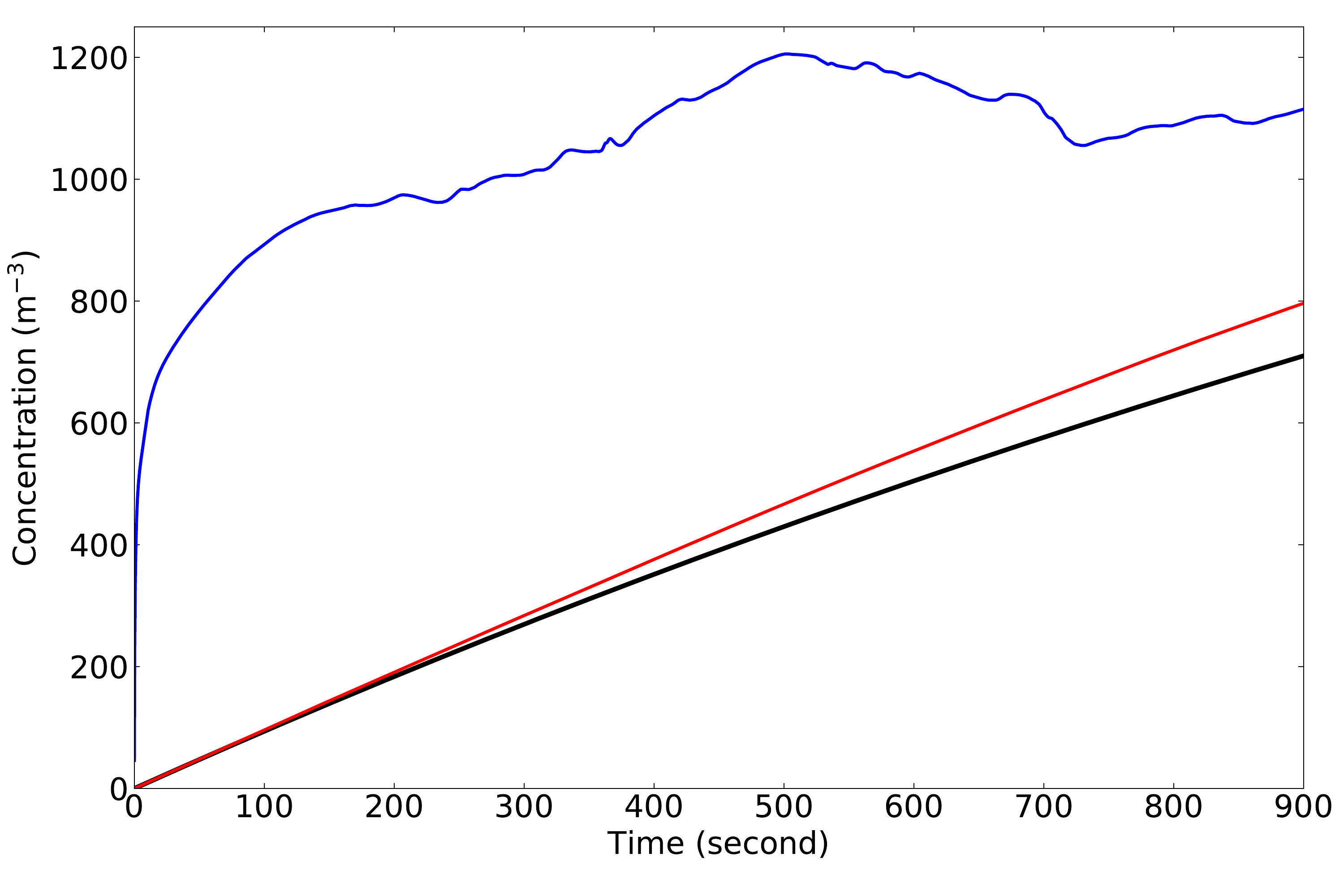}
        \caption{Run 3} \label{fig:average-run3}
    \end{subfigure}
    \hspace*{\fill}
    \begin{subfigure}{0.48\textwidth}
        \includegraphics[width=\linewidth]{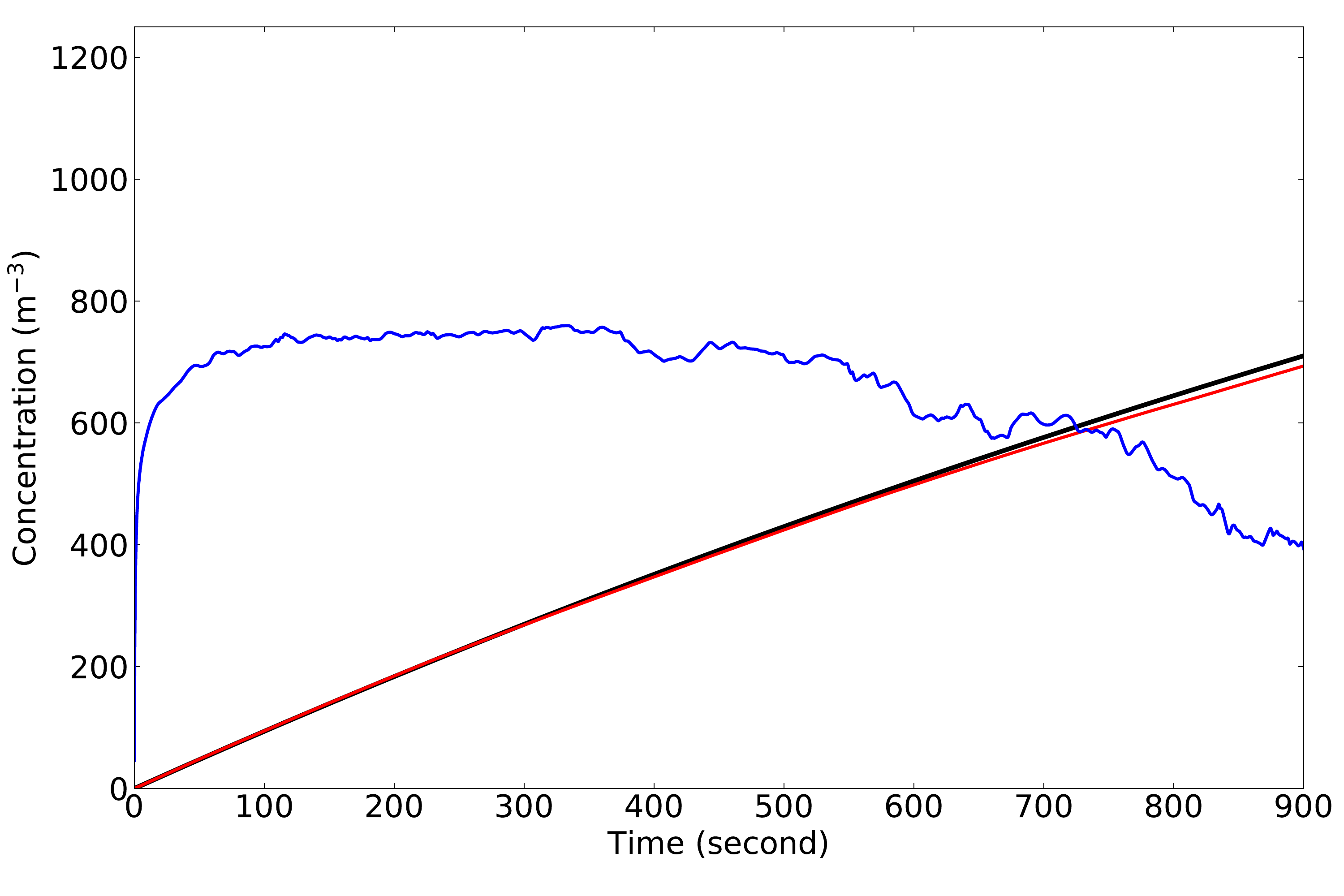}
        \caption{Run 4} \label{fig:average-run4}
    \end{subfigure}
    \caption{Time history of average and standard deviation of concentration.}
    \label{fig:average-C}
\end{figure}

\begin{figure}
    \centering
    \begin{subfigure}{0.9\textwidth}
        \caption{Run 1} \label{fig:pdf-run1}
        \includegraphics[width=\linewidth]{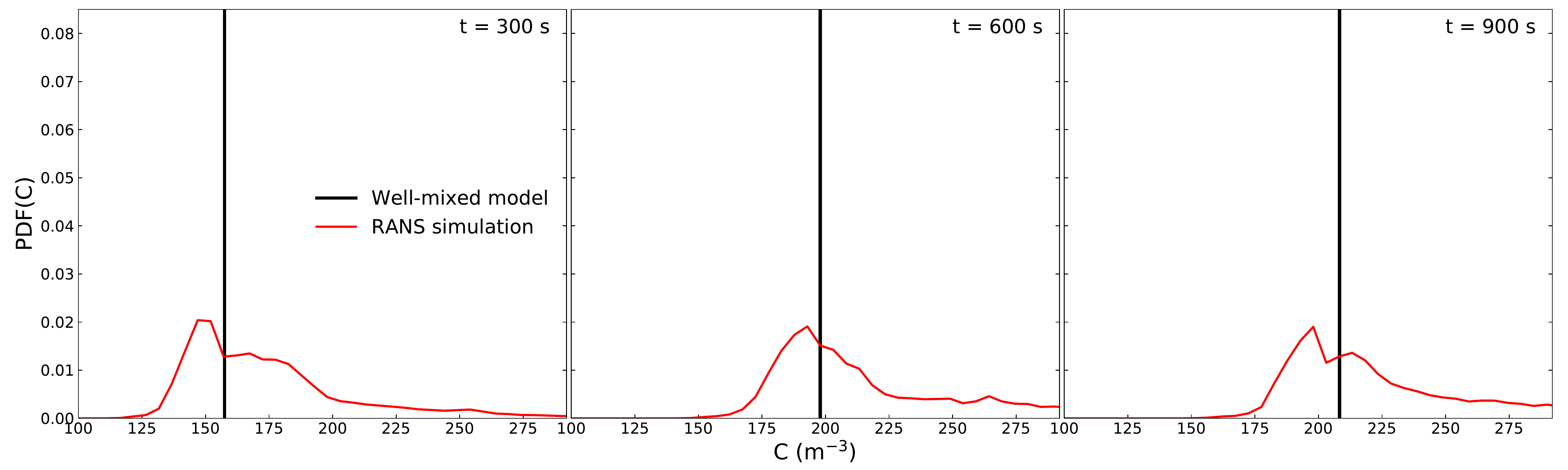}
    \end{subfigure}
    \centering
    \begin{subfigure}{0.9\textwidth}
        \caption{Run 2} \label{fig:pdf-run2}
        \includegraphics[width=\linewidth]{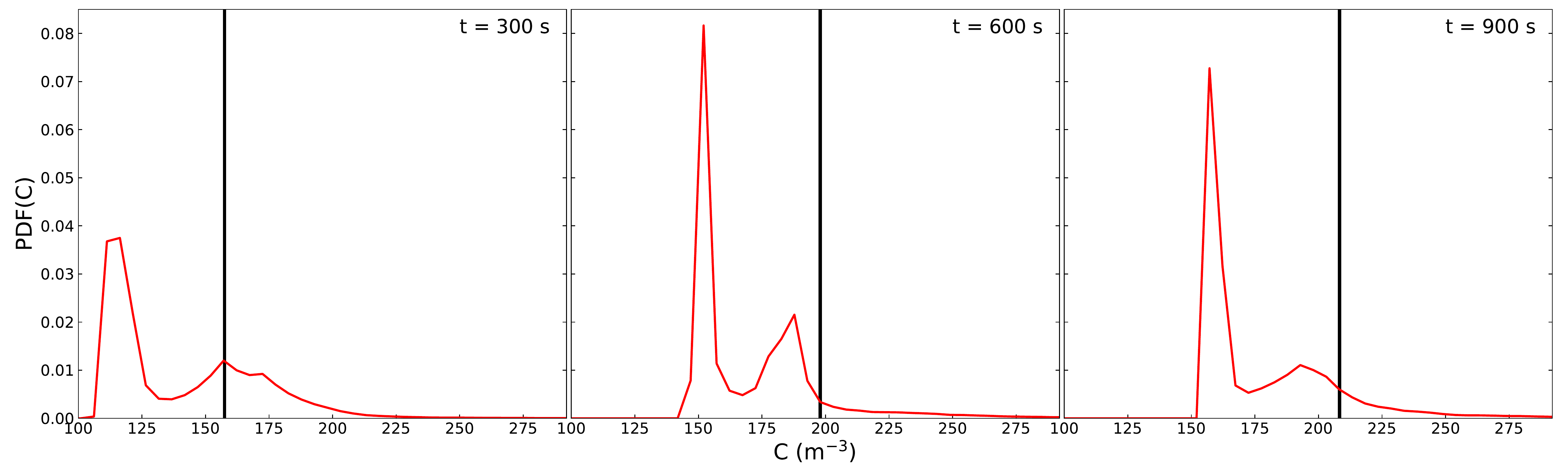}
    \end{subfigure}
    \centering
    \begin{subfigure}{0.9\textwidth}
        \caption{Run 3} \label{fig:pdf-run3}
        \includegraphics[width=\linewidth]{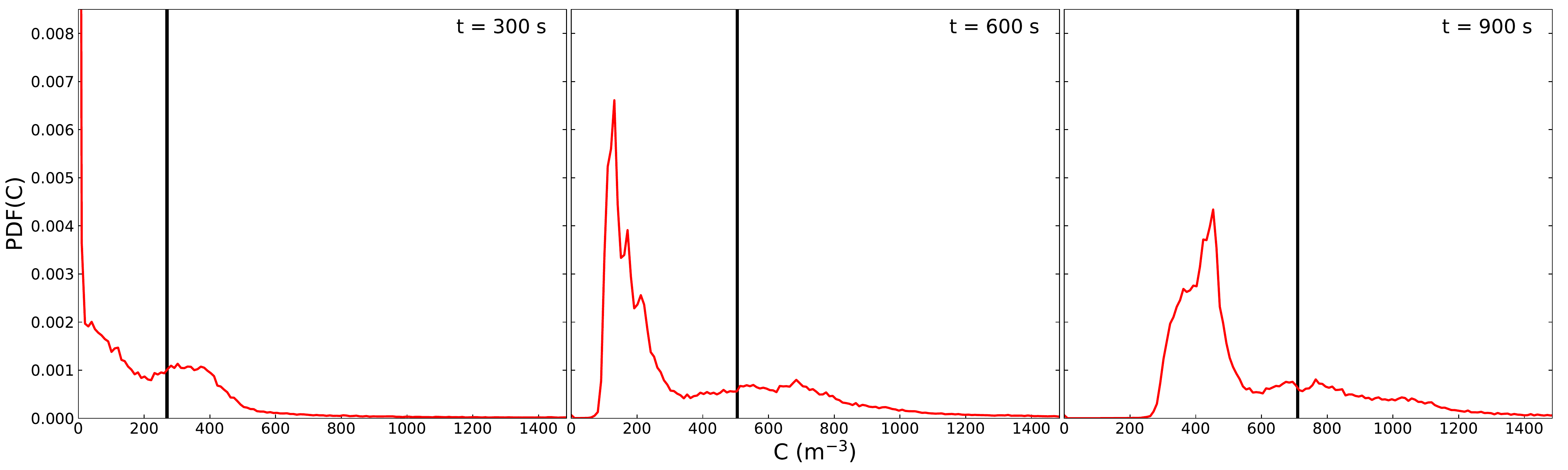}
    \end{subfigure}
    \centering
    \begin{subfigure}{0.9\textwidth}
        \caption{Run 4} \label{fig:pdf-run4}
        \includegraphics[width=\linewidth]{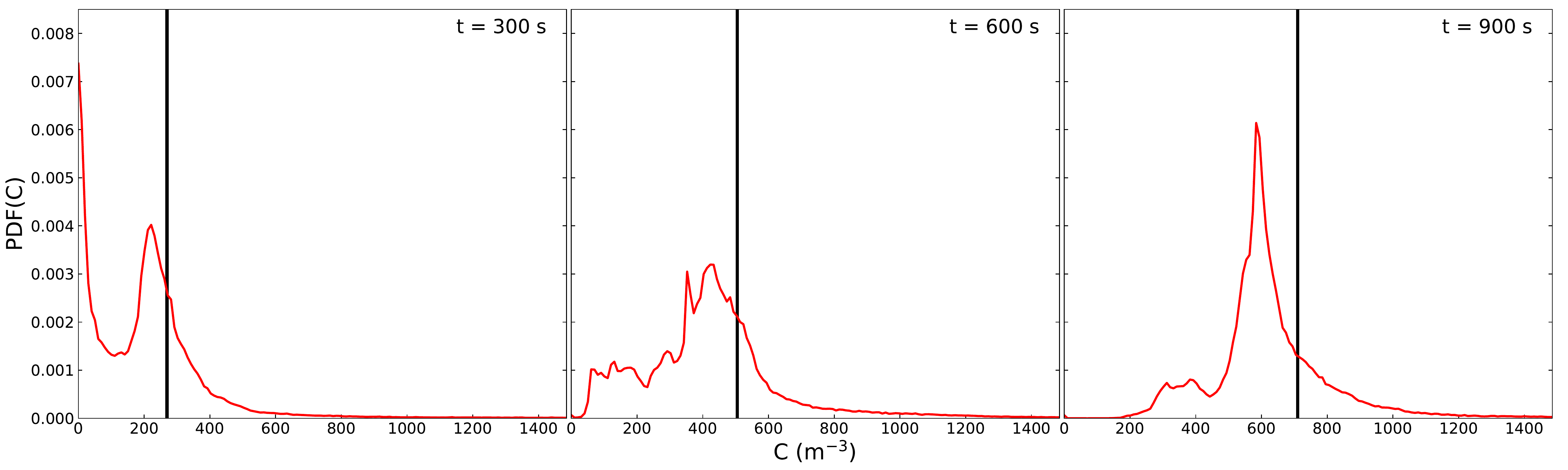}
    \end{subfigure}
    \caption{PDF of concentration at different time instants.}
    \label{fig:pdf}
\end{figure}

The mean and standard deviation of the concentration are plotted as a function of time in Fig.~\ref{fig:average-C}, compared with the well-mixed model. Although the well-mixed model gives close average values, the large values of standard deviation indicate that the spatial distribution exhibits strong non-uniformity.  This is more clearly shown by the Probability Density Function (PDF) of concentration in Fig.~\ref{fig:pdf}, which samples over the entire domain. The well-mixed model assumes uniform distribution hence has only a single value at each time, while the distribution from the RANS simulations is spread out over several orders of magnitude. For each run, more than one peak can be observed in the PDF. This may indicate different processes that influence the concentration field, such as diffusion and convection. The peaks shown by the simulations are to the left of the value from the well-mixed model, which means the well-mixed model overpredicts the concentration over the majority of space whereas there exits pockets of high concentration that are represented by the long tails of the PDF.

\subsection{Inhaled Aerosols}

\begin{figure}[!ht]
     \includegraphics[width=0.5\linewidth]{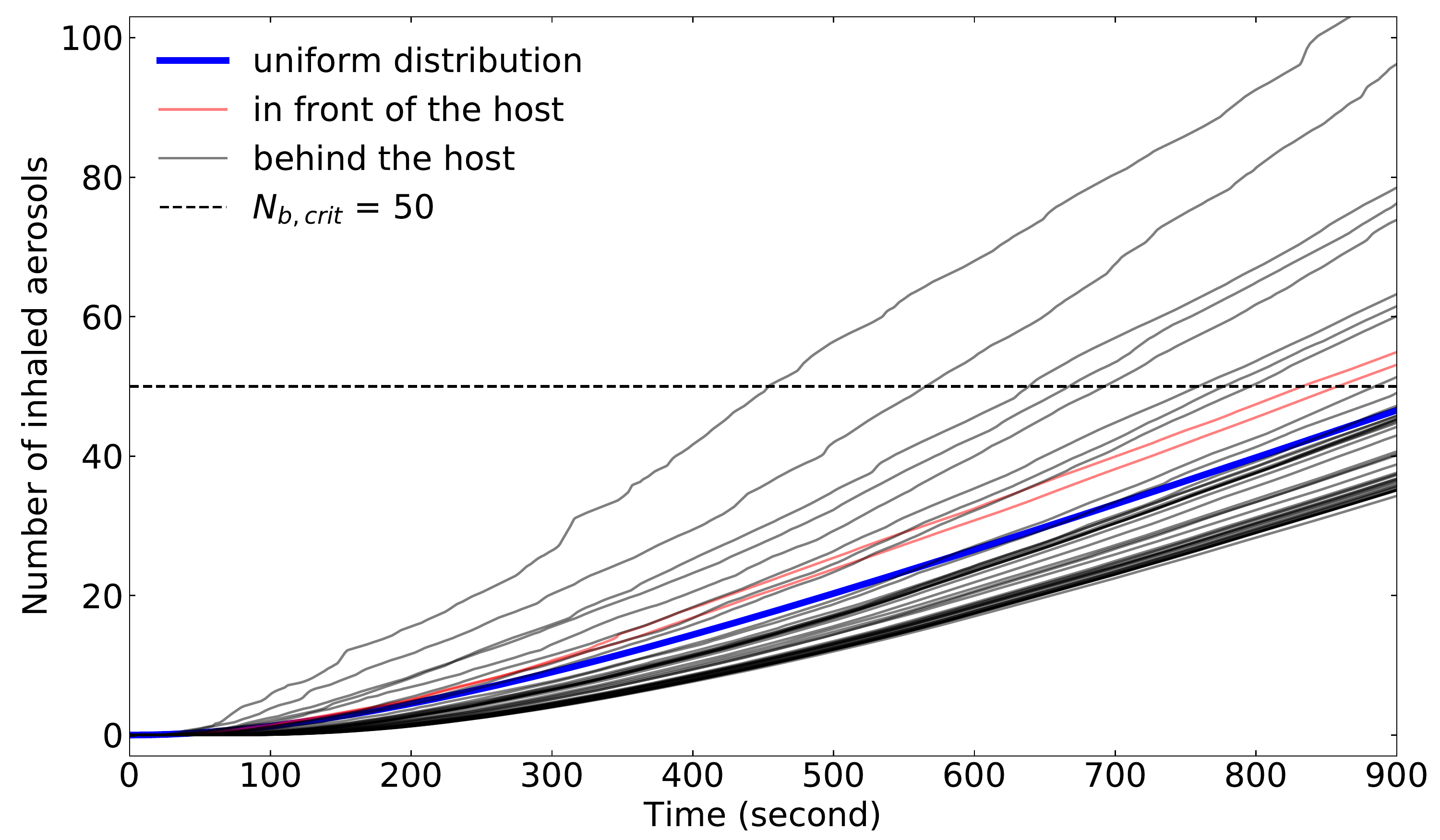}

     \includegraphics[width=0.5\linewidth]{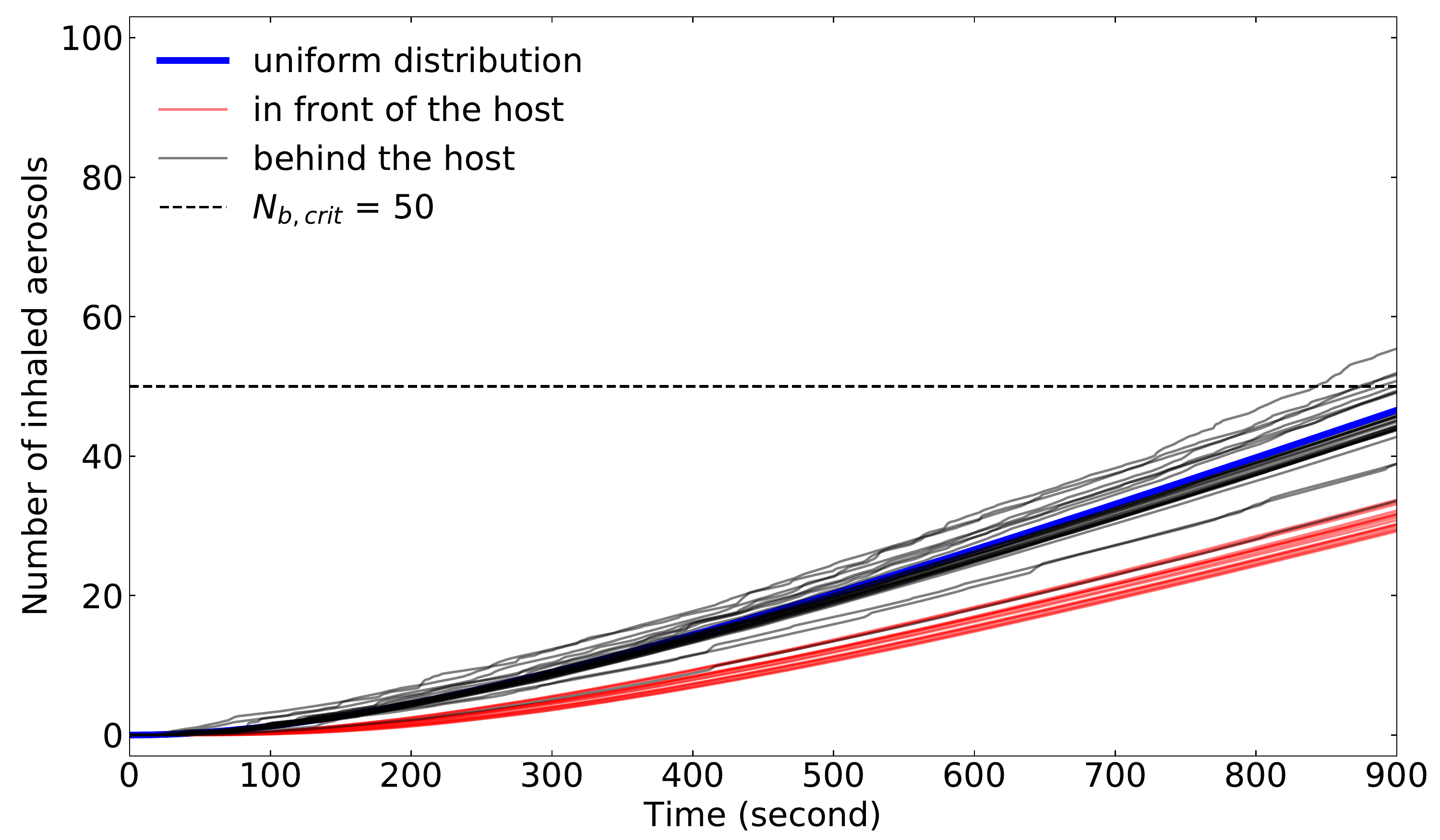}
     
     \includegraphics[width=0.5\linewidth]{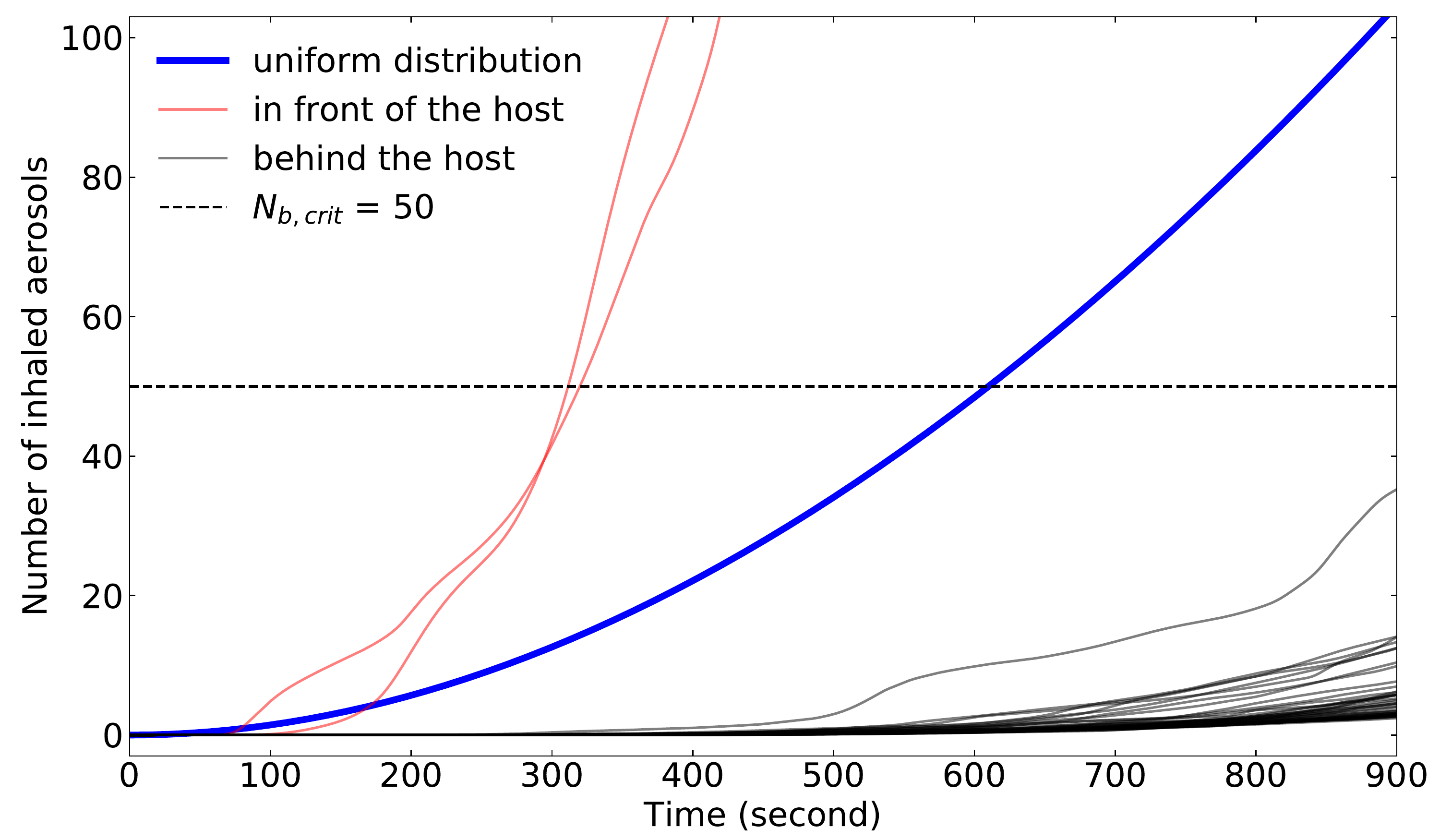}

     \includegraphics[width=0.5\linewidth]{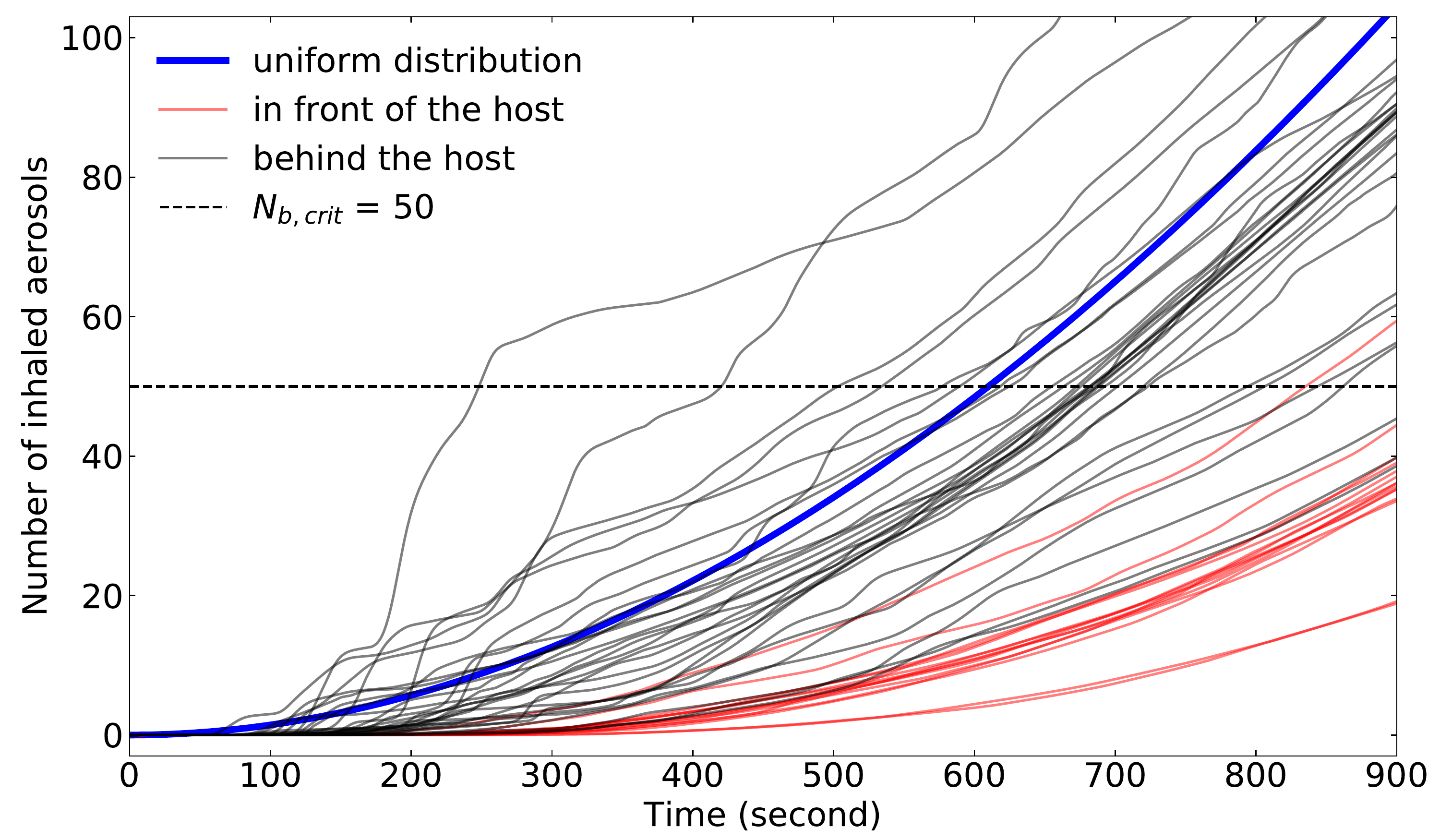}
     \caption{Time history of number of inhaled aerosols at different passenger locations, compared with the well-mixed model (blue lines). Top to bottom are Runs 1--4.}
     \label{fig:inhaled}
\end{figure}

The number of inhaled aerosols is determined through numerical solution of Eq.~\eqref{eqn:n}.  Figure~\ref{fig:inhaled} shows the time history of $N_b$ for each passenger, together with the solution of Eq.~\eqref{eqn:n-well-mixed} that assumes the aerosols are well-mixed. The CFD results are colored according to the relative position in the bus.

Figure~\ref{fig:inhaled} shows the number of inhaled aerosols increases faster than linearly for the entire 15~min exposure window.  Some of the passengers in the front of the bus are exposed to higher concentrations, and are the first to exceed the MID threshold in as soon as  450~s. For the passengers to the rear of the bus, the concentration predicted by the CFD is less than that of the well-mixed model, and they do not exceed the MID in the 15~min ride.

Run~2 shows similar behavior as Run~1. The high HVAC rate dominates the airflow, the turbulence mixes the aerosol well, and the agreement between the well-mixed model and the CFD is strong.

For Run~3, the HVAC rate is reduced, which localizes the aerosol field. Most passengers inhale fewer aerosols, with the exception of two passengers near the host.  The two that exceed the MID do so in the first 300~s on the bus, whereas the most of the others inhale no more than 15 aerosols on the entire ride.

Figure~\ref{fig:inhaled} shows the result for Run~4.  A hallmark of the low HVAC rate is the large fluctuation in the velocity and concentration fields, and hence the number of inhaled aerosols. Here, many of the passengers inhale more than the MID, the first of which after only 200~s on the bus.

\begin{figure}[!ht]
	\centering
	\begin{subfigure}{1\textwidth}
		\caption{Run 1}
		\includegraphics[width=0.95\textwidth]{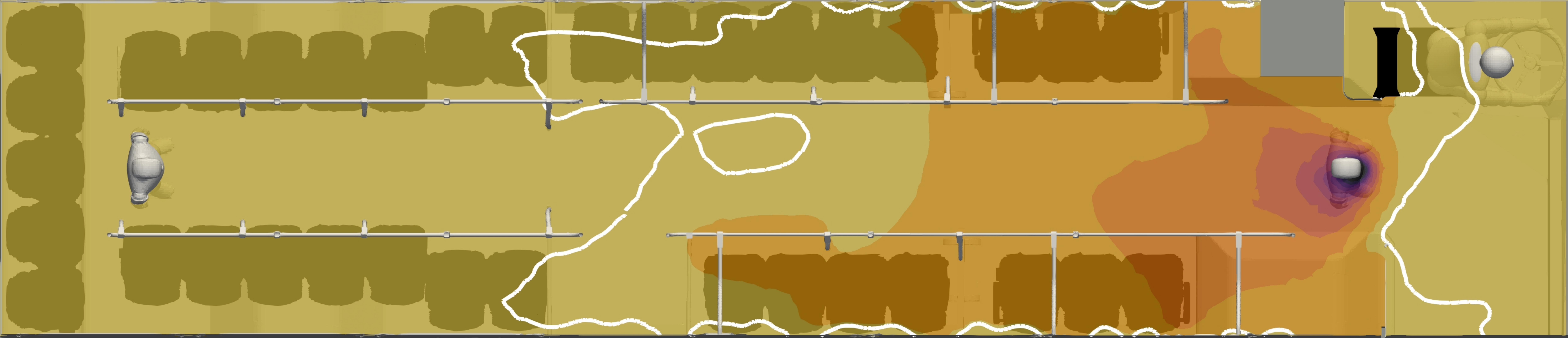}
	\end{subfigure}

	\begin{subfigure}{1\textwidth}
		\caption{Run 2}
		\includegraphics[width=0.95\textwidth]{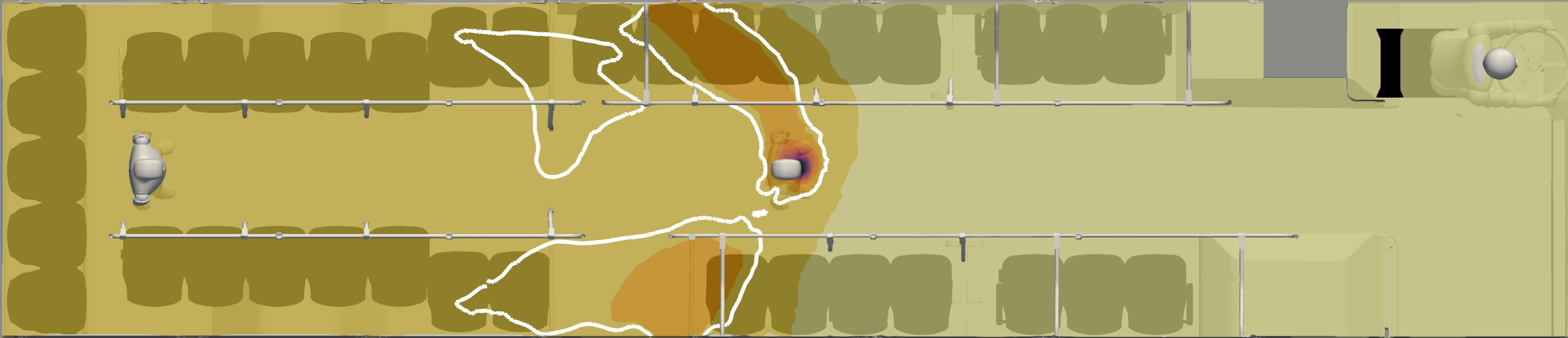}
	\end{subfigure}
		\begin{subfigure}{1\textwidth}
		\caption{Run 3}
		\includegraphics[width=0.95\textwidth]{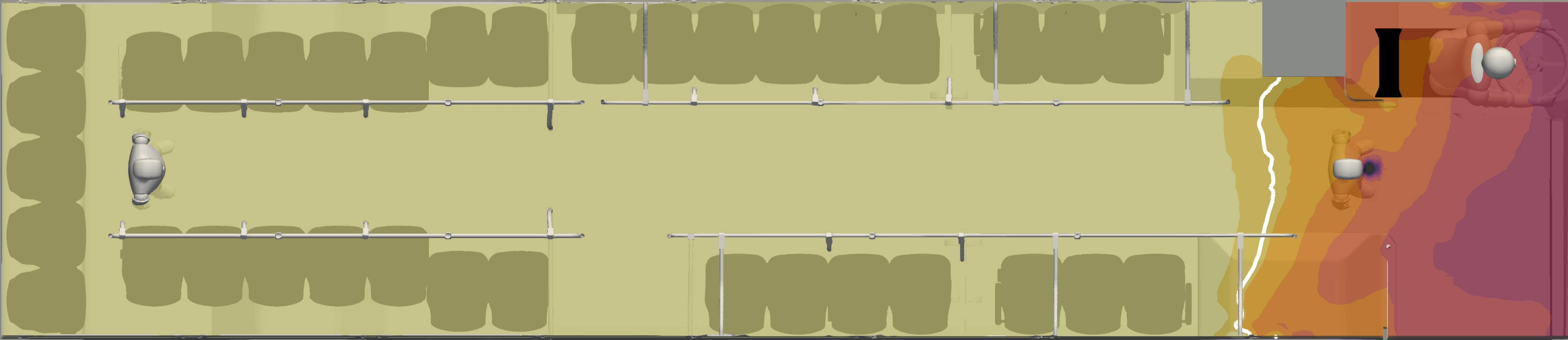}
	\end{subfigure}
	\begin{subfigure}{1\textwidth}
		\caption{Run 4}
		\includegraphics[width=0.95\textwidth]{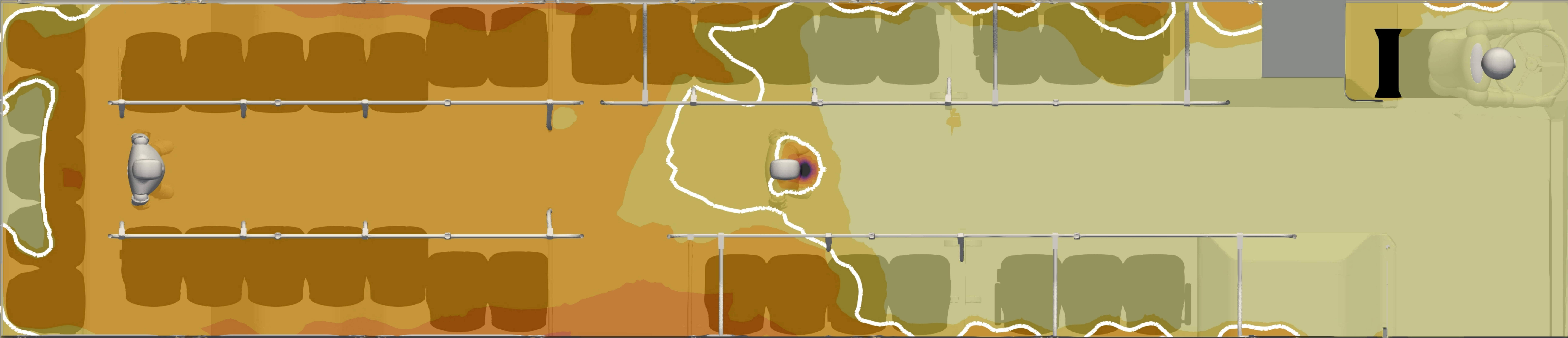}
	\end{subfigure}
	\begin{subfigure}{0.7\textwidth}
        \includegraphics[width=\linewidth]{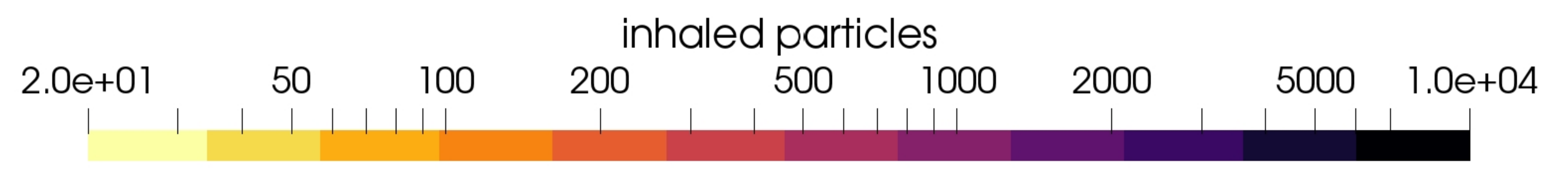}   
    \end{subfigure}
\caption{\label{fig:contours} Contours of inhaled aerosols at $t=15$ min. The white contour lines represent the critical number of inhaled aerosols $N_{\rm b,crit} = 50$.}
\end{figure}

The exposure to the virus in the bus can be visualized by the contour of the quantity~$N_b$.  This is shown in Fig.~\ref{fig:contours} for the plane of 1.7~m above the lower deck at $t=15$~min. The contours are spaced logarithmically, and the contour of the MID is shown in white. 

\subsection{Number of Infected Passengers}

Guidance on how to safely operate busses can be summarized by calculating the number of susceptible passengers that become infected under different conditions.  In this section, this quantity is determined by counting the number of passengers that inhale more than the MID. Comparison is also made with the Wells--Riley~\citep{Riley78} prediction of number of infected passengers (Eqn.~\ref{eqn:WR}).  This equation uses the well-mixed assumption to predict the equilibrium concentration. Also, we introduce a simple mask model based on \cite{Mittal20b} to assess their effectiveness while being worn on the bus. 

\begin{figure}[!ht]
    \centering
    \begin{subfigure}{0.48\textwidth}
        \includegraphics[width=\linewidth]{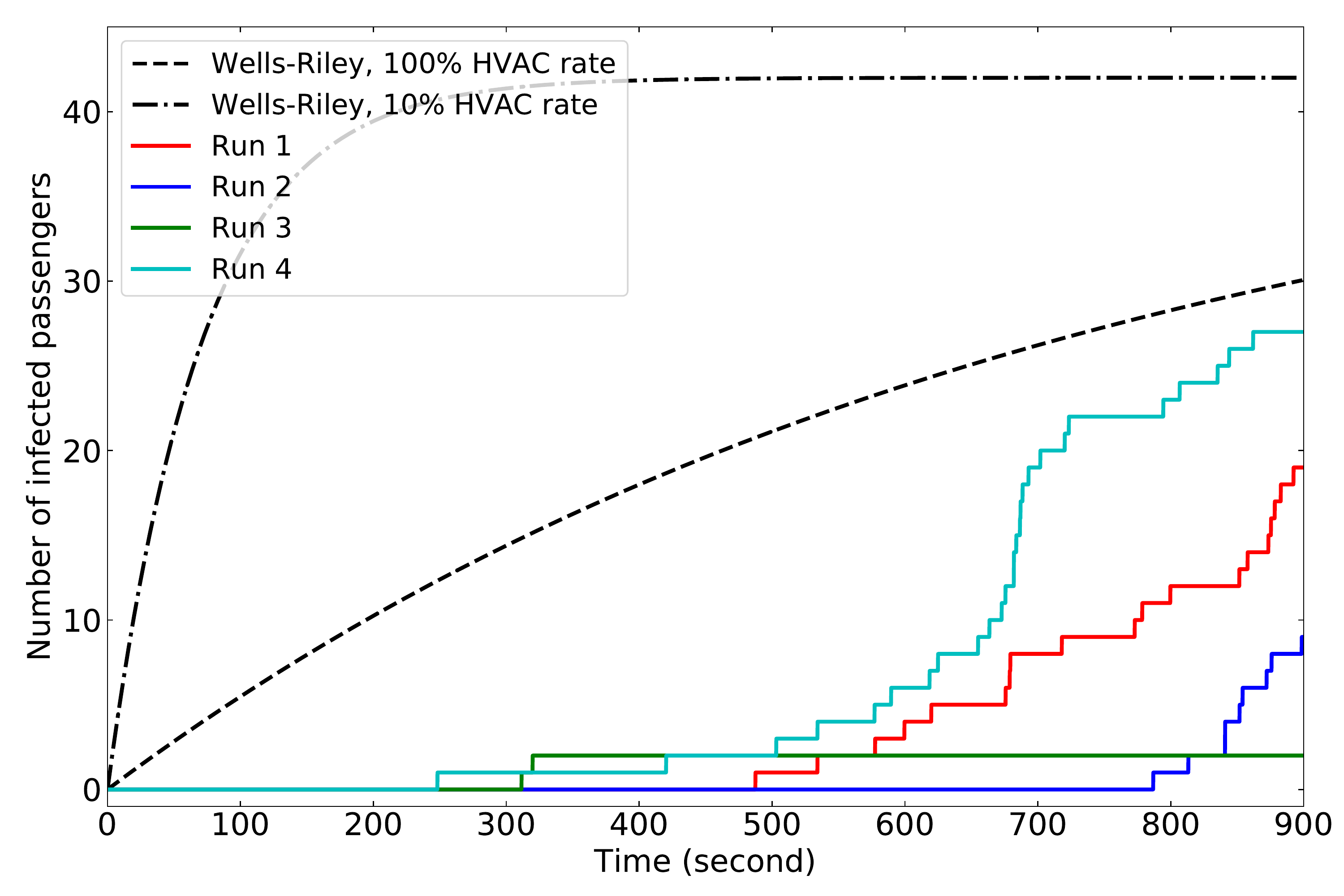}
        \caption{no mask}
     	\label{fig:num-infected-noMask}
    \end{subfigure}
    \hspace*{\fill}
    \begin{subfigure}{0.48\textwidth}
        \includegraphics[width=\linewidth]{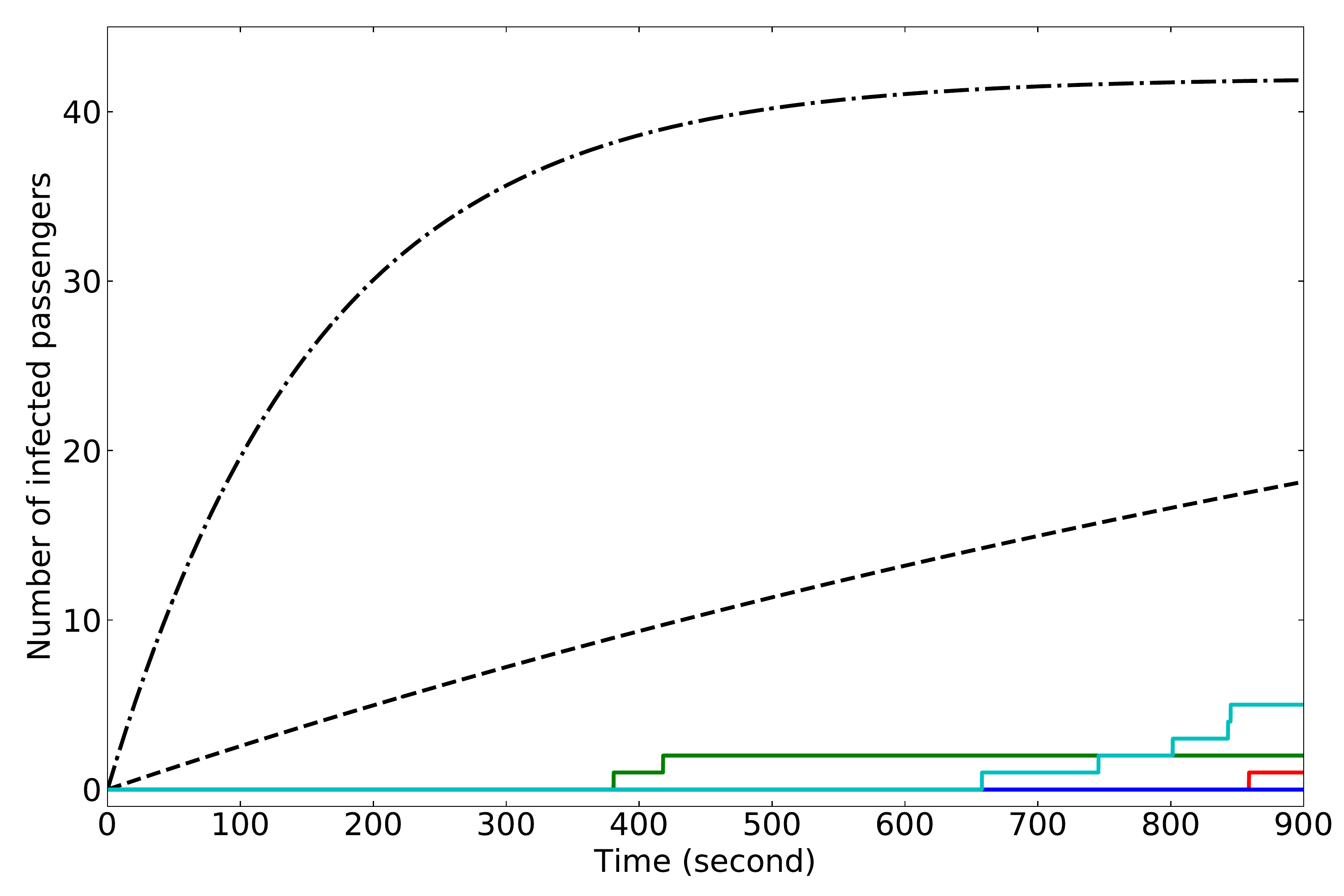}
        \caption{everyone wears a handmade mask.}
     \label{fig:num-infected-DIY}
    \end{subfigure}
    \caption{Number of infected passengers as a function of time, compared with Wells--Riley model.}
    \label{fig:num-infected}
\end{figure}

Figure~\ref{fig:num-infected} shows the number of passengers that inhale more than MID for each run.  Note that the Wells--Riley prediction for 100\% HVAC rate corresponds to Runs~1 and 2, and the 10\% HVAC rate curve corresponds to Runs 3 and 4.  There are a total of 42 susceptible passengers.   
This figure highlights the complexity of the transmission process.  Starting from Run 1, there is a rapidly increasing number of infections, reaching the value of 19 by the end of the ride.  If the infected person is in the middle of the bus (Run 2), the number of infected decreases by 50\%, whereas the well-mixed assumption and Wells--Riley model do not distinguish the effect of infected passenger location.  For this specific HVAC design, conditioned air is supplied uniformly along the roof of the bus, but the return is only in the rear.  This sets up a net rearward convection velocity field, which excludes passengers in the front of the bus from inhaling infectious particles (other than those that are mixed with fresh air and recirculated to the passenger compartment).

Run 3 shows that only 2 passengers become ill, yet 27 contract the disease in Run 4. For both these runs the HVAC rate is minimal, and thus less forced mixing by the mechanical ventilation system. For both of these runs, the plume of high-concentration remains close to the infected passenger.  When they are in the front of the bus, few other passengers are within range, whereas when they are in the middle of the bus, the high concentration field encompasses more than half the passengers.

\paragraph{Simple mask model}

The previous results are all based on the assumption that the host emits aerosols at a high rate. A key strategy to combat the spread of the disease is to wear a face covering or mask, which has the two-fold effect of reducing the number of aerosols emitted from the host, and reducing the number that are inhaled by the susceptible passenger.  The physics of masks are complex for many reasons and are the subject of intense scientific study.  In order to qualitatively assess the influence of masks,  the results are reanalyzed using a simple model based on  \citep{Dbouk20, Mittal20b}.  The mask model assumes that a fraction of both the emitted and inhaled aerosols are blocked by the mask. Since mask effectiveness depends on many factors, including the type of mask and nature of how it is worn, two values of mask effectiveness are investigated.  To represent a well-fitted surgical mask it is assumed that 90\% of the aerosols are blocked, and to more conservatively estimate the effectiveness of masks a lower effectiveness of 30\% is also investigated.

When the 90\% effective mask is worn by all passengers, there are no transmissions predicted by either the CFD or the well-mixed model.

Figure~\ref{fig:num-infected-DIY} shows the number of infected passengers as a function of time when each person wears a handmade mask with effectiveness of 30\%.  The CFD simulations show that for masks are effective, the reduction in transmissions is significant when compared to Fig.~\ref{fig:num-infected}.  Also, the difference between the Wells--Riley model and the CFD is significant, where the model shows all passengers will become infected at $t\approx880$~s for the lower HCAC rate.

\section{Summary}

In this paper, the transient nature of aerosol dispersion within an urban bus is studied with CFD and a well-mixed model. Two different mechanical ventilation settings are analyzed corresponding to strong and weak mechanical ventilation.   The spatial mean and standard deviation of the concentration are computed as a function of time, and for all cases the standard deviation is significantly larger than the mean, at least for the initial time.  

For the high HVAC setting, the concentration throughout the cabin follows the general trend that is predicted by the well-mixed model, although the model either over or underpredicts the CFD data depending on location of the host.  The number of infected aerosols is also calculated, and the trends of the CFD and well-mixed model are similar to each other, but again depending on the location of the host and susceptible passenger the well-mixed model either over or underpredicts number of inhaled aerosols.  This means that while the well-mixed model should be used with caution, and its accuracy depends on many factors including relative position of the infectious individual and the airflow in the enclosed space.

Finally, a minimum infective dose is used to count the number of infected passengers as a function of time and compared to the Wells--Riley model. The differences between the detailed simulations and the well-mixed model are significant.  The number of infected passengers grows in time as the aerosols travel throughout the passenger compartment, whereas the Wells--Riley model assumes uniform concentration throughout and significantly overpredicts the number of transmissions.

At low HVAC rate, the results are differ even more.  In the absence of mechanical ventilation, mixing is done via buoyancy and diffusion. This leads to more localized concentration of the aerosol around the host, and consequently the well-mixed model is a worse approximation of the concentration in the bus.  Interesting, when the infected person is in the front, the CFD predicts the fewest number of transmissions among the four cases analyzed herein.  At the same low HVAC setting but with the infected person in the middle, the CFD predicts the largest number of transmissions.  As the mechanical ventilation rate becomes lower, the well-mixed model becomes a worse indicator of the concentration in the bus for the passengers, this is true when considering them individually, and as a whole.

Finally, a mask model is implemented assess their role on mitigation.  When a mask with 90\% effectiveness is worn by all passengers, there are no transmissions on the bus for all cases considered in the paper. When a 30\% effective mask is worn by all, the number of transmission is significantly reduced compared to when no masks are worn, although there are significant differences between the CFD and the Wells--Riley equation.

In summary, the air flow inside the cabin of an urban bus is complex and plays a critical role on the transmissibility of airborne diseases. In some circumstances, especially at higher HVAC rates, a simple well-mixed model does a reasonable job predicting the trends in mean concentration. However, it is unable to capture the variation about the mean, which the CFD results demonstrated to be significant. Incorporating spatial information into a simple model is the focus of future work.

\begin{acknowledgments}
We wish to acknowledge the support of the University of Michigan College of Engineering and the Advanced Research Computing Technology Services for providing the computational resources to perform the simulations.
\end{acknowledgments}

\nocite{*}
\bibliography{myRef.bib}

\end{document}